\documentclass[twocolumn,trackchanges]{aastex63}

\newcommand{\Msun}{$M_{\odot}$}
\newcommand{\Rsun}{$R_{\odot}$}
\newcommand{\Rstar}{R$_{\ast}$}
\newcommand{\RE}{R$_{\Earth}$}

\shorttitle{Unresolved Binary Exoplanet Host Stars Fit as Single Stars}
\shortauthors{Furlan \& Howell}
\submitjournal{ApJ; accepted on June 10, 2020}

\begin{document}

\title{Unresolved Binary Exoplanet Host Stars Fit as Single Stars: Effects on the Stellar Parameters}

\author[0000-0001-9800-6248]{E. Furlan}
\affiliation{NASA Exoplanet Science Institute, Caltech/IPAC, Mail Code 100-22, 1200 E. California Blvd., 
Pasadena, CA 91125, USA}

\author[0000-0002-2532-2853]{S. B. Howell}
\affiliation{NASA Ames Research Center, Moffett Field, CA 94035, USA}

\correspondingauthor{E. Furlan}
\email{furlan@ipac.caltech.edu}

\begin{abstract}
In this work we quantify the effect of an unresolved companion star on the derived 
stellar parameters of the primary star if a blended spectrum is fit assuming the star 
is single. Fitting tools that determine stellar parameters from spectra typically fit for 
a single star, but we know that up to half of all exoplanet host stars may have one or 
more companion stars. We use high-resolution spectra of planet host stars 
in the {\it Kepler} field from the California-{\it Kepler} Survey to create simulated 
binaries; we select 8 stellar pairs and vary the contribution of the secondary star, then 
determine stellar parameters with {\tt SpecMatch-Emp} and compare them to the
parameters derived for the primary star alone. 
We find that in most cases the effective temperature, surface gravity, metallicity, 
and stellar radius derived from the composite spectrum are within 2-3 $\sigma$
of the values determined from the unblended spectrum, but the deviations depend 
on the properties of the two stars. Relatively bright companion stars that are similar 
to the primary star have the largest effect on the derived parameters; in these cases 
the stellar radii can be overestimated by up to 60\%. We find that metallicities are 
generally underestimated, with values up to 8 times smaller than the typical 
uncertainty in [Fe/H]. Our study shows that follow-up observations are necessary 
to detect or set limits on stellar companions of planetary host stars so that stellar 
(and planet) parameters are as accurate as possible.
\end{abstract}


\section{Introduction}
\label{intro}

The discovery of an exoplanet orbiting its host star is the beginning of a process 
that aims at culminating with the determination of detailed properties of both the 
star and the planet. Only then can additional characterization work, such as transit 
spectroscopy or the assessment of potential habitability, be fully achieved. 
Over the last two decades, several missions and surveys have discovered many 
hundreds of exoplanets, most notable the over 4,700 confirmed planets and planet
candidates discovered by the {\it Kepler} mission \citep{borucki16}. Follow-up 
observations using space- and ground-based telescopes have provided imaging and 
spectroscopic details for the exoplanet-hosting stars \citep[e.g.,][]{howell11, 
adams12, adams13, lillo-box12, lillo-box14, dressing14, horch14, law14, marcy14, 
wang14, wang15a, wang15b, cartier15, everett15, gilliland15, torres15, baranec16, 
kraus16, ziegler17, ziegler18, furlan17, furlan18}. Detailed follow-up continues 
for the planet candidates found by the {\it Transiting Exoplanet Survey Satellite} 
({\it TESS}; \citealt{ricker15}), relying on both TESS team members as well as the 
community.

We have learned that the phrase “Know thy star, know thy planet” rings true, as 
the more accurately the host star parameters are known, the more definitively
we can characterize any exoplanets it harbors. Transit observations, such as the ones 
carried out by {\it Kepler} and {\it TESS}, give us, in addition to some of the planet's
orbital parameters, the exoplanet radius, but it depends on the radius of the star it orbits 
and whether the photometric aperture is contaminated by ``third light'', i.e.,
unresolved stellar companions. Such a companion will dilute the transit, causing us to 
observe a shallower transit depth and leading us to derive a smaller planet radius \citep{ciardi15}. 
\citet{horch14} and \citet{matson18} have shown that approximately half of all exoplanet host 
stars may be binaries (or higher-order multiples), similar to the binary fraction observed in field
stars \citep{raghavan10}. Other studies have shown that certain (mostly close) binaries are 
less likely to be planet host stars \citep{wang14,kraus16}. Moreover, about 30\% of 
binaries in the solar neighborhood are comprised of about equal-mass stars (mass ratio $>$ 0.8);
for the closest binaries ($<$ 100 au), this fraction increases to $\sim$ 40\% \citep{raghavan10}.
Such bright companions ($\Delta$m $\lesssim$ 1 for a Sun-like star), which would have the 
largest transit dilution effect, are thus fairly common.
It is now well-accepted that in order to determine the correct transit depth, one must correct 
for the third light, that is, even perfect knowledge of the stellar radius is not enough. 
Without correction, the exoplanet radius and mean density will be incorrect 
\citep{furlan17,hirsch17,furlan17b,teske18}, and calculations of the atmospheric scale 
height will be skewed \citep{batalha17}. In fact, this situation can be even more 
insidious as it is not always clear which star the exoplanet actually transits. 

Corrections for a companion star (bound or line of sight) are now commonplace to adjust 
transit depths \citep[e.g.,][]{howell19}, but such corrections are not generally applied to 
spectral observations. Nearby, unresolved companions could result in incorrect stellar 
parameters, given that their spectral features are added to those of the primary 
star, but tools that extract stellar parameters from spectra typically fit for a single star 
\citep[e.g.,][]{torres12,endl16,petigura17}. 

\citet{kolbl15} did some work related to this idea by searching the large database of 
high-resolution spectra of the California Planet Search (CPS; \citealt{howard10}) for
blended companion stars. They refit the high-resolution spectra of host stars of {\it Kepler}
planet candidates, seeking to determine if a single star was most appropriate or if, in
addition, a second star could be fit. They searched for spectral signatures of a
companion star in the residual spectrum, after the best fit to the primary star had been
subtracted. They found spectral evidence for companions in 63 sources (out of a 
sample of 1160 stars). 
\citet{teske15}, using high-resolution imaging data, made an attempt to confirm the 
suspected companions in order to provide matches in these two techniques, allowing 
a better understanding of host systems. Unfortunately, no cases agreed between the two 
studies, leaving the situation of spectral decomposition and direct imaging confirmation 
a bit confused. The method of \citet{kolbl15} could be affected by incomplete line 
lists and certain differences in radial velocity and luminosity between the primary
and companion star that make the detection of the companion unfeasible. However,
this also suggests that the two techniques generally probe different populations of 
binaries, and also different binary separations, with just a small overlap in parameter 
space. 

Stellar properties are determined accurately from high-resolution spectroscopy
\citep[e.g.,][]{torres12,mortier13,petigura17,johnson17}; the most commonly used 
techniques rely on model stellar atmospheres and atomic and molecular line lists. 
Some methods apply synthetic models to fit the spectral lines; these models are either 
pre-calculated to form a library \citep[e.g.,][]{buchhave12, petigura15,endl16} or are 
synthesized to achieve a best fit to the observed spectrum \citep[e.g.,][]{valenti96,valenti05}. 
Another method uses equivalent widths of Fe I and Fe II lines and compares them to 
model widths derived from model atmospheres, assuming LTE and excitation and 
ionization equilibrium \citep[e.g.,][]{sneden73,santos04,mortier13,teixeira16}. 
For transiting planet host stars, the stellar densities can be determined directly from 
the transit light curve; through isochrone fits, the surface gravity of the star can then 
be derived \citep{sozzetti07}.

The stellar parameters of effective temperature ($T_{\mathrm{eff}}$), surface gravity 
($\log (g)$), and metallicity ([Fe/H]) are the observational values obtained, while fitting
using stellar evolution models (such as the Dartmouth Stellar Evolution Program 
isochrones; \citealt{dotter08}) and/or asteroseismology leads to additional parameters 
such as radius, mass and age.
Combined with parallaxes from {\it Gaia}, stellar radii can be determined with 
a precision of $\sim$3-8\% \citep[][for the {\it Kepler} sample]{berger18,fulton18}. 
Stellar radius and mass further constrain exoplanet parameters such as the planet 
radius and, combined with the orbital period, insolation flux and habitable zone inclusion.
In addition, uncertainties in the stellar parameters, both from the data and the 
models used to derive them, contribute to the uncertainties of the planet parameters 
from the transit fit (for the planet radius) and radial velocity fit (for the planet mass).

Using large samples, general statistical methods can be applied, as in \citet{huber13}, to 
refine global stellar properties and check on overall correctness. For example, \citet{huber13}
took the pursuit of stellar parameters to a highly refined level by combining spectroscopic 
observations and asteroseismology to determine stellar radii to $\pm$3\% and stellar 
masses to $\pm$7\%. Compared to just spectroscopically derived stellar parameters, 
which are more precise than those inferred from photometry and have typical uncertainties 
of $\sim$~15\% in stellar radius and $\sim$~10\% in stellar mass \citep{torres12, muirhead12, 
mortier13, huber14, johnson17, mathur17}, these are very accurate values and lead 
to more definitive exoplanet parameters. 
However, these types of analysis assume that the host star is single; if that is 
not the case, a spatially close companion star may make it more difficult to measure stellar
oscillations \citep{sekaran19} and can produce spectral contamination that will lessen the 
accuracy of the stellar fitting procedures’ final values. Oscillations can also be changed 
by tidal interactions in close binaries, from being suppressed \citep{schonhut20} to being 
excited at specific frequencies \citep{fuller17}.

Approximately half the stars are not single: for solar-type stars within 25 pc of the Sun, 
about 45\% have at least one companion star \citep{raghavan10}. For these nearby, 
multiple stellar systems, 11\% of companions have periods less than 1000 days; this 
fraction increases to 21\% and 40\% for periods less than 10$^4$ and 10$^5$ days,
respectively \citep{raghavan10}. Assuming a combined mass of 1.5 \Msun, a period of 
10$^4$ days corresponds to a semi-major axis of $\sim$ 10 AU; projected on the sky, 
10 AU is less than 1\arcsec\ at distances beyond 10 pc, and so the binary system would 
likely be unresolved in seeing-limited imaging and spectroscopy. Considering that the peak
of the period distribution of companions in the \citet{raghavan10} sample lies between 
10$^3$ and 10$^6$ days, and many exoplanet host stars lie at distances of a few 
hundred pc, we expect most bound companions to exoplanet host stars to be 
unresolved in spectra obtained with $\sim$ 1\arcsec-wide slits.

We seek to quantify the amount and type of additional error an undetected stellar 
companion might cause in the spectral fitting determination of stellar parameters. 
We have used a few high-resolution spectra of {\it Kepler} planet host stars, created 
blends, and fit these simulated binaries with {\tt SpecMatch-Emp} \citep{yee17}. 
We provide quantitative and qualitative estimates of how stellar blends affect the stellar 
parameters derived from these simulated spectra of unresolved stellar systems. For faint 
companions, the contamination is small and can be ignored. But companions that 
are at least half as bright as the primary star lead to a complex contamination matrix 
of their influence on the determined stellar properties of $T_{\mathrm{eff}}$, $\log (g)$, 
and [Fe/H], resulting in values of these stellar parameters than can deviate up to 
2-3 $\sigma$ from the values derived from an unblended spectrum (where $\sigma$
is the uncertainty returned by the fitting code). In turn, these unaccounted-for 
deviations lessen the accuracy of the parameters determined for any orbiting exoplanet. 

We describe the selection of the spectra analyzed for this study and our methodology
to create blended spectra and derive their stellar parameters in section \ref{sample}, 
the results of our stellar fits in section~\ref{results}, discuss the implications in 
section~\ref{discuss}, and give our conclusions in section~\ref{summ}.

\vspace{3ex}

\section{Sample and Methodology}
\label{sample}

We selected 16 stars from the California-{\it Kepler} Survey (CKS) to combine their spectra 
and create blended systems. As part of the CKS, 1305 stars in the {\it Kepler} field were 
observed with HIRES \citep{vogt94} on the Keck I telescope, with the goal of determining 
more accurate stellar parameters for {\it Kepler} planet host stars and thus for their transiting 
planets \citep{petigura17,johnson17}.
As described in \citet{petigura17}, two fitting codes, {\tt SpecMatch} and {\tt SME@X-SEDE},
were used to derive stellar parameters from the HIRES spectra of the 1305 stars in the CKS 
sample. The former code, which was actually developed for the CKS project to analyze 
Keck/HIRES spectra \citep{petigura15}, interpolates between model spectra to fit an 
observed spectrum. The latter code, which is based on Spectroscopy Made Easy (SME;
\citealt{valenti96}), calculates synthetic spectra via a radiative transfer code applied to model 
atmospheres. The stellar effective temperatures, surface gravities, and metallicities derived 
with these two methods agree very well, typically within the measurement uncertainties 
\citep{petigura17,brewer18}. Their combined values were incorporated into a 
catalog\footnote{Both the catalog and HIRES spectra from the CKS are available at 
https://california-planet-search.github.io/cks-website/}.

\begin{deluxetable*}{lcccccc} \scriptsize
\tablewidth{0.8\linewidth}
\tablecaption{Stellar Parameters Derived with {\tt SpecMatch-Emp} for the Eight Stellar Pairs
Used in This Work to Create Simulated Binaries
\label{star_params}}
\tablehead{
\colhead{KOI} & \colhead{$T_{\mathrm{eff}}$ [K] } & \colhead{$\log (g)$} & \colhead{[Fe/H]} & 
\colhead{$R_{\ast}$ [\Rsun]} & \colhead{$\chi^2$} & \colhead{SNR}}
\startdata
\multicolumn{7}{c}{{\it Binary 1} ($\Delta T_{\mathrm{eff}}=$1404 K, $L_{\mathrm{sec}}/L_{\mathrm{prim}}=$0.15)} \\
2711 & 5882 $\pm$ 110 & 4.42 $\pm$  0.12 &  0.08 $\pm$ 0.09 & 1.06 $\pm$ 0.18 & 1.95 & 53.0 \\
448 & 4478 $\pm$   70 & 4.62 $\pm$  0.12 &  0.04 $\pm$ 0.09 & 0.70 $\pm$ 0.10 & 6.51 & 29.8 \\ 
\hline
\multicolumn{7}{c}{{\it Binary 2} ($\Delta T_{\mathrm{eff}}=$1218 K, $L_{\mathrm{sec}}/L_{\mathrm{prim}}=$0.19)} \\
692 & 5664 $\pm$  110 & 4.25 $\pm$  0.12 &  0.19 $\pm$ 0.09 & 1.06 $\pm$ 0.18 & 1.86 & 50.9 \\  
1871 & 4446 $\pm$   70 &  4.60 $\pm$  0.12 &  0.20 $\pm$ 0.09 & 0.74 $\pm$ 0.10 & 7.96 & 25.8 \\ 
\hline
\multicolumn{7}{c}{{\it Binary 3} ($\Delta T_{\mathrm{eff}}=$874 K, $L_{\mathrm{sec}}/L_{\mathrm{prim}}=$0.30)} \\
2559 & 5791 $\pm$  110 &  4.29 $\pm$  0.12 &  0.10 $\pm$ 0.09 & 1.02 $\pm$ 0.18 & 2.18 & 48.4 \\ 
757 & 4917 $\pm$  110 &  4.53 $\pm$  0.12 &  0.10 $\pm$ 0.09 & 0.77 $\pm$ 0.10 & 8.59 & 24.4 \\  
\hline
\multicolumn{7}{c}{{\it Binary 4} ($\Delta T_{\mathrm{eff}}=$722 K, $L_{\mathrm{sec}}/L_{\mathrm{prim}}=$0.43)} \\
5622 & 5272 $\pm$  110 &  4.55 $\pm$  0.12 &  0.05 $\pm$ 0.09 & 0.82 $\pm$ 0.10 & 8.86 & 23.8 \\ 
870 & 4550 $\pm$  110 &  4.57 $\pm$  0.12 &  0.09 $\pm$ 0.09 & 0.72 $\pm$ 0.10 & 12.04 & 20.9 \\ 
\hline
\multicolumn{7}{c}{{\it Binary 5} ($\Delta T_{\mathrm{eff}}=$363 K, $L_{\mathrm{sec}}/L_{\mathrm{prim}}=$0.51)} \\
1089 & 5736 $\pm$  110 &  4.27 $\pm$  0.12 &  0.08 $\pm$ 0.09 & 1.03 $\pm$ 0.18 & 7.10 & 28.0 \\ 
749 & 5373 $\pm$  110 &  4.53 $\pm$  0.12 &  0.05 $\pm$ 0.09 & 0.84 $\pm$ 0.10 & 3.73 & 37.6 \\ 
\hline
\multicolumn{7}{c}{{\it Binary 6} ($\Delta T_{\mathrm{eff}}=$403 K, $L_{\mathrm{sec}}/L_{\mathrm{prim}}=$0.61)} \\
869 & 4989 $\pm$  110 &  4.52 $\pm$  0.12 &  0.13 $\pm$ 0.09 & 0.79 $\pm$ 0.10 & 11.50 & 21.4 \\ 
2339 & 4586 $\pm$  110 &  4.54 $\pm$  0.12 &  0.15 $\pm$ 0.09 & 0.73 $\pm$ 0.10 & 13.04 & 21.1 \\  
\hline
\multicolumn{7}{c}{{\it Binary 7} ($\Delta T_{\mathrm{eff}}=$132 K, $L_{\mathrm{sec}}/L_{\mathrm{prim}}=$0.66)} \\
116 & 5892 $\pm$  110 &  4.37 $\pm$  0.12 & -0.11 $\pm$ 0.09 & 1.12 $\pm$ 0.18 & 0.43 & 135.3 \\ 
1379 & 5760 $\pm$  110 &  4.49 $\pm$  0.12 & -0.11 $\pm$ 0.09 & 0.95 $\pm$ 0.10 & 2.90 & 45.4 \\ 
\hline
\multicolumn{7}{c}{{\it Binary 8} ($\Delta T_{\mathrm{eff}}=$32 K, $L_{\mathrm{sec}}/L_{\mathrm{prim}}=$0.84)} \\
4072 & 5816 $\pm$  110 &  4.22 $\pm$  0.12 &  0.12 $\pm$  0.09 & 1.20 $\pm$  0.18 & 1.97 & 53.0 \\  
3422 & 5784 $\pm$  110 &  4.42 $\pm$  0.12 &  0.12 $\pm$  0.09 & 1.11 $\pm$  0.18 & 1.70 & 57.3 \\  
\enddata
\tablecomments{The first column lists the Kepler Object of Interest (KOI) number of the star. 
All stars host at least one planet candidate, with most stars (all except for KOI 1871, 2559, 
and 3422) hosting confirmed planets. 
The $\chi^2$ value is the median of the $\chi^2$ values returned by {\tt SpecMatch-Emp} for
each 100 {\AA} segment of the 5000$-$5800 {\AA} spectrum. The last column lists an estimate
of the signal-to-noise ratio of the spectrum.}
\end{deluxetable*}

From the catalog of stellar parameters derived by the CKS team, we selected 16 targets 
to create blended spectra (see Table \ref{star_params}). 
In order to simulate spectra of unresolved binary stars consisting of a G-type star and a 
cooler companion, we selected dwarf stars with $T_{\mathrm{eff}}$ in the 5000-6000 K 
range for the ``primary'' star and dwarf stars with $T_{\mathrm{eff}}$ around 4400-5000 K 
range for the ``secondary'' star (with three exceptions, for which the companion star was 
chosen to be $<$ 400 K cooler than the primary star). In addition, we also selected pairs 
of stars with similar metallicities (matching within 0.1 dex) to mimic binaries formed out 
of the same molecular cloud core and thus with the same initial composition. 
Multiplicity surveys of solar-type stars located within a few tens of parsecs from the Sun 
show that there is a roughly flat distribution in the mass ratios of the secondary and
primary stars, with a small deficit for the lowest-mass companions (mass ratios $\lesssim$ 0.2)
and an overabundance of about equal-mass companions \citep{raghavan10, tokovinin14}.
So, our stellar pairs mimic binaries that are actually observed.

As a next step, we used the reduced HIRES spectra of our selected 16 stars from the 
CKS sample and determined their stellar parameters ($T_{\mathrm{eff}}$, $\log (g)$, 
[Fe/H], and $R_{\ast}$) using {\tt SpecMatch-Emp}\footnote{https://github.com/samuelyeewl/specmatch-emp}.
This fitting code is a different version of {\tt SpecMatch} \citep{yee17}. {\tt SpecMatch-Emp} uses 
a library of observed spectra of calibrator stars to determine stellar parameters; the use of an 
``empirical'' library results in more accurate fits for mid- to late-K and M stars (which are more 
difficult to fit with synthetic spectra, given numerous atomic and molecular lines with poorly 
known properties). The library contains 404 stars that were observed with Keck/HIRES as 
part of the California Planet Search (CPS; \citealt{howard10}); these library stars have spectra
with sufficiently high signal-to-noise (most have S/N $>$ 100) and well-determined stellar parameters 
from spectroscopy, spectrophotometry, interferometry, and asteroseismology \citep{yee17}. 
These parameters were retrieved from the literature (see \citealt{yee17} for details); for
stars without a complete set ($T_{\mathrm{eff}}$,  $\log (g)$, [Fe/H], $R_{\ast}$, $M_{\ast}$),
\citet{yee17} derived the missing parameters by fitting to the Dartmouth grid of stellar models
\citep{dotter08}. The stellar parameters of the library stars cover $\sim$ 3000$-$7000 K in 
$T_{\mathrm{eff}}$, $\sim$ 0.1$-$16 {\Rsun} in $R_{\ast}$, and $-0.6$ to $+0.6$ dex in [Fe/H].
The uncertainties returned by {\tt SpecMatch-Emp} are set by the scatter of the differences 
between the stellar parameters derived by {\tt SpecMatch-Emp} for the library stars and 
their library values. These uncertainties are smaller for cool stars, since their library 
parameters are more accurate.

Given that {\tt SpecMatch-Emp} uses the 5000$-$5800 {\AA} range for its spectral fits, we only 
used the central ('r') HIRES spectrum (which covers 4975$-$6420 {\AA}).
When fitting a spectrum with {\tt SpecMatch-Emp}, it is first shifted onto the library wavelength
scale (to account for the line-of-sight velocity of the target), then ``matched'' to the library spectra 
to find the best-matching spectra (which includes line broadening and normalization), and finally 
the parameters from the five best-matching spectra are combined by a weighted average to 
determine the stellar parameters \citep[see][]{yee17}. To find the best-matching 
library spectra and then their best-fitting linear combination, an unnormalized ${\chi}^2$ statistic 
is used; to account for differences in continuum normalization, a cubic spline is fit to the residuals 
as ${\chi}^2$ is minimized. For both the matching and combination steps, the spectra are 
divided into 100 {\AA} segments, and only wavelengths between 5000 and 5800 {\AA} are 
used (a smaller range could be used, too, but it might result in less reliable stellar parameters).

\begin{figure*}[!t]
\centering
\includegraphics[scale=0.52]{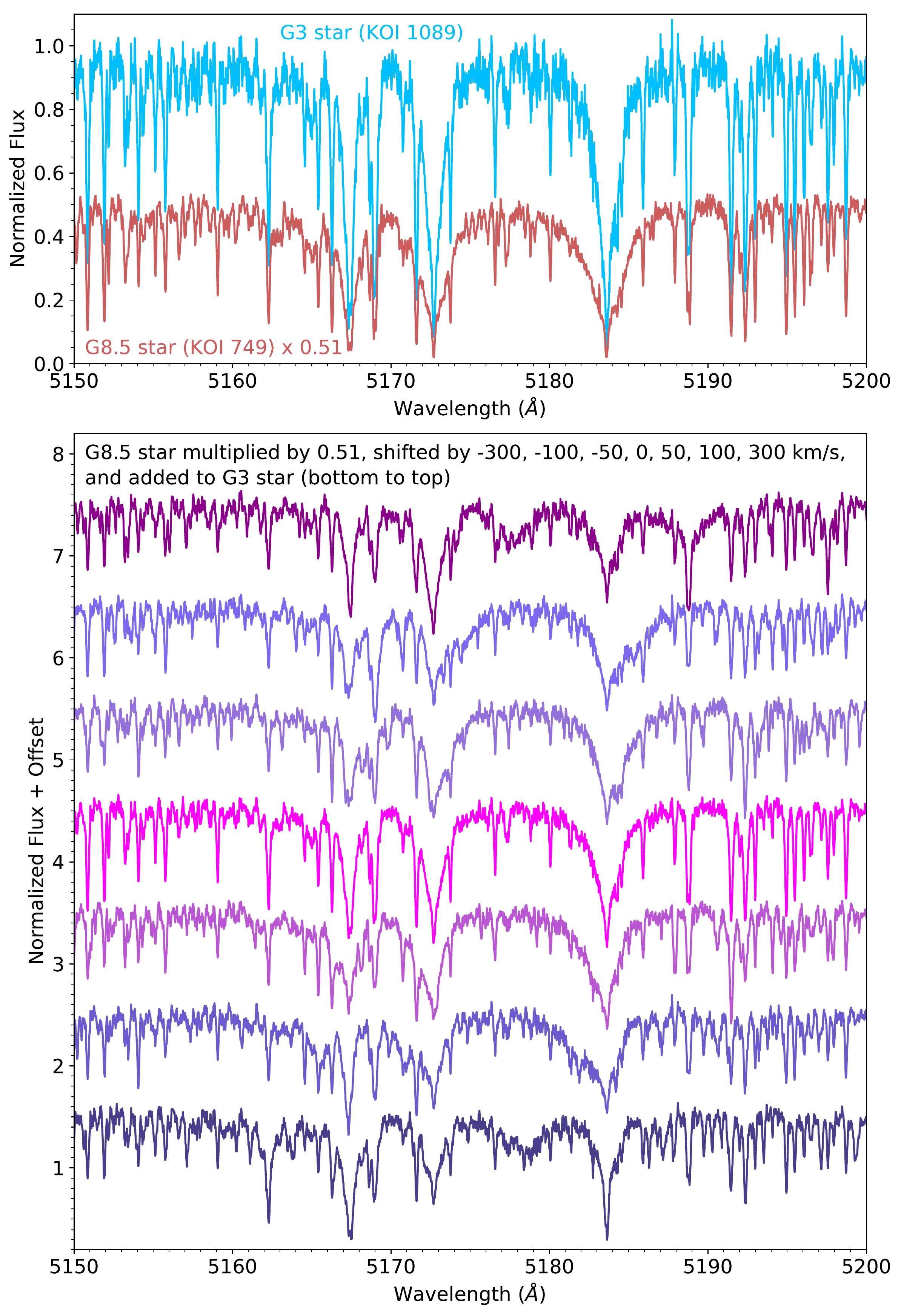}
\caption{Combination of two CKS spectra to create a simulated binary (only the 5150--5200 {\AA}
region is shown). The top panel displays the two spectra, where the spectrum of the cooler star 
has been multiplied by a factor equal to the luminosity ratio of the two stars. The bottom panel 
shows representative combinations of the two spectra after the multiplicative factor and RV shifts 
have been applied to the cooler star.
\label{binary_example}}
\end{figure*}

We derived the stellar parameters for our 16 targets using {\tt SpecMatch-Emp} (see Table 
\ref{star_params}). In most cases the derived stellar parameters agreed within the uncertainties 
with the parameters from the CKS catalog. Deviations of up to 3$\sigma$ in $T_{\mathrm{eff}}$ 
were found for the coolest stars ($T_{\mathrm{eff}}$ $\sim$ 4400-4600 K), with the values 
derived with {\tt SpecMatch-Emp} smaller by $\sim$ 120-180 K. This may be expected, given 
that {\tt SpecMatch-Emp} is more accurate for stars with $T_{\mathrm{eff}}$ $\lesssim$ 4500 K 
\citep{yee17}. For two of the cool stars (KOI 448 and 870), \citet{muirhead12} used a 
different method to determine effective temperatures by measuring spectral indices derived 
from $K$-band spectra and obtained even lower values (by 500-600 K). However, this method 
becomes more uncertain for values larger than $\sim$ 3800 K (see \citealt{muirhead12} for 
details), and at least for KOI 870 the formal uncertainties imply that the derived $T_{\mathrm{eff}}$ 
value agrees within 1$\sigma$ with the value derived with {\tt SpecMatch-Emp}. Therefore, 
the $T_{\mathrm{eff}}$ uncertainties for these cooler stars could be somewhat larger than 
their formally derived uncertainties, but likely not more than a factor of two. 
In this work, we focus on the stellar parameters of the ``primary'' stars, which all have 
$T_{\mathrm{eff}}$ values in the 5000-5900 K range, so even if the stellar parameters 
of cooler stars are in some cases more unreliable, they should not significantly 
affect the results of our fits. Moreover, we use the same fitting code for all spectra, so 
our results are self-consistent and allow sensible comparisons.

We created simulated binary stars by combining the spectra of the pairs listed in
Table \ref{star_params}. We started with the ``shifted'' version of the spectra, after 
 {\tt SpecMatch-Emp} has shifted the spectra onto the library wavelength scale.
These eight pairs range from a difference in $T_{\mathrm{eff}}$ of 1404 K to just 
32 K and a ratio in luminosity of the secondary relative to the primary star from 
0.15 to 0.84. 
Given that the reduced HIRES spectra are normalized, the spectrum of the companion 
star should be multiplied by a factor less than 1 before being added to the spectrum
of the primary star in order to simulate a realistic binary companion. 
The luminosity ratio represents a rough approximation of the flux ratio in the optical 
(given differences in the spectral type and thus spectral energy distribution of our stellar 
pairs, we expect the optical flux ratios to be smaller for binaries with the coolest
secondaries, ranging from about 0.6 to 0.7 of the luminosity ratios). 
We therefore scaled the companion stars by the luminosity ratio calculated from 
the effective temperatures and stellar radii given in Table \ref{star_params} so they 
resemble bound secondaries. 

Moreover, the signal-to-noise ratio (SNR) of the spectra varies (see Table
\ref{star_params}), but in a bound system, where the combined spectrum is obtained,
the contribution of the brighter star should have a larger SNR than that of the fainter star.
Thus, before adding the two spectra, we degraded the SNR of one of the spectra by adding
Gaussian noise to make sure that the ratio of the two SNRs is roughly equal to the square 
root of the luminosity ratio (as a proxy for the brightness ratio in the optical). In most cases,
the SNR was degraded by less than 30\% and so had just a minor effect on the spectrum.
As noted by \citet{yee17}, the {\tt SpecMatch-Emp} algorithm is quite robust even at low
SNR (as low as 10); its accuracy is more limited by the matching process than the noise
in the spectrum.  

After scaling and adjusting the SNR of the spectra, we applied different radial velocity 
(RV) shifts to the spectrum of the companion star, ranging from $-500$ to $+500$ km 
s$^{-1}$, to represent probable orbital motion of the two stars around the center of mass. 
Depending on the inclination angle of the binary system's orbit, the eccentricity and
orbital period, and the mass ratio of the two stars, the semi-amplitude of the RVs of 
the secondary star ($K_2$) can vary from just a few km s$^{-1}$ to several hundred 
km s$^{-1}$. For example, we estimated that for binaries on circular orbits seen at 
intermediate inclination angles ($\sim$ 60\degr) and with orbital periods between 
100 and 3000 days, the $K_2$ values range from $\sim$ 10 to 60 km s$^{-1}$ 
(assuming mass ratios between 0.1 and 1.0); on orbits with shorter periods, $K_2$ 
values can reach over 250 km s$^{-1}$, with even higher values for more edge-on
orientations and small secondary masses. These $K_2$ values are also in
agreement with observations \citep[e.g.,][]{tokovinin18}. Thus, we expect 
the majority of $K_2$ values in a broad distribution over the 0--100 km s$^{-1}$ 
range, with fewer values between 100 and 500 km s$^{-1}$. Based on this motivation,
we chose RV shifts of $\pm 500$, $\pm 400$, $\pm 300$, $\pm 200$, $\pm 100$, 
$\pm 90$, $\pm 80$, $\pm 70$, $\pm 60$, $\pm 50$, $\pm 40$, $\pm 30$, $\pm 20$, 
$\pm 10$, and also 0 km s$^{-1}$ for the companions star's spectrum.
The range of RV shifts we probed here are representative for close binaries, for 
which individual (i.e., spatially resolved) spectra are difficult to impossible to obtain.

Examples of the process of creating simulated binaries are shown in Figure 
\ref{binary_example}, where we display the two individual spectra of a pair 
(after they have been shifted to the same wavelength frame and the cooler star
scaled by the luminosity ratio of the two stars) and then various representative
combinations of the two spectra by applying an RV shift to the scaled spectrum of 
the cooler star before addition to the hotter star's spectrum.

\begin{deluxetable*}{llcccccc} \scriptsize
\tablewidth{0.8\linewidth}
\tablecaption{Stellar Parameters Derived with {\tt SpecMatch-Emp} from the Blended Spectra
of the the 232 Simulated Binaries
\label{star_params_binaries}}
\tablehead{
\colhead{Star 1} & \colhead{Star 2} & \colhead{RV [km s$^{-1}$]} & \colhead{$T_{\mathrm{eff}}$ [K] } 
& \colhead{$\log (g)$} & \colhead{[Fe/H]} & \colhead{$R_{\ast}$ [\Rsun]} & \colhead{$\chi^2$}}
\startdata
\multicolumn{7}{c}{{\it Binary 1} ($\Delta T_{\mathrm{eff}}=$1404 K, $L_{\mathrm{sec}}/L_{\mathrm{prim}}=$0.15)} \\
KOI2711  & KOI0448  &   -500  &   5956 $\pm$ 110 &  4.39   $\pm$ 0.12 &   -0.00  $\pm$ 0.09 &    1.11 $\pm$ 0.18 &   4.584 \\
KOI2711  & KOI0448  &   -400  &   5966 $\pm$ 110 &    4.38   $\pm$ 0.12 &   -0.02  $\pm$ 0.09 &    1.12 $\pm$ 0.18 &  4.407 \\ 
KOI2711  & KOI0448  &   -300  &   5986  $\pm$ 110 &    4.41  $\pm$ 0.12 &    0.00  $\pm$ 0.09 &    1.09 $\pm$ 0.18 &  4.312 \\
KOI2711  & KOI0448  &   -200  &   6000 $\pm$ 110 &    4.40  $\pm$ 0.12 &    0.02  $\pm$ 0.09 &    1.13 $\pm$ 0.18 &  4.351 \\
KOI2711  & KOI0448  &   -100  &   5968 $\pm$ 110 &    4.40  $\pm$ 0.12 &   -0.00  $\pm$ 0.09 &    1.11 $\pm$ 0.18 &  4.662 \\
KOI2711  & KOI0448  &    -90   &  5955 $\pm$ 110 &    4.38  $\pm$ 0.12 &   -0.02  $\pm$ 0.09 &    1.13 $\pm$ 0.18 &  4.526  \\
KOI2711  & KOI0448  &    -80   &  5970 $\pm$ 110 &    4.42  $\pm$ 0.12 &   -0.00  $\pm$ 0.09 &    1.08 $\pm$ 0.18 &  4.528 \\ 
KOI2711  & KOI0448  &    -70   &  5959 $\pm$ 110 &    4.42  $\pm$ 0.12 &   -0.01  $\pm$ 0.09 &    1.08 $\pm$ 0.18 &  4.580 \\ 
KOI2711  & KOI0448  &    -60   &  5965 $\pm$ 110 &    4.43  $\pm$ 0.12 &   -0.00  $\pm$ 0.09 &    1.08 $\pm$ 0.18 &  4.588 \\ 
KOI2711  & KOI0448  &    -50   &  5961 $\pm$ 110 &    4.44  $\pm$ 0.12 &    0.01  $\pm$ 0.09 &    1.07 $\pm$ 0.18 &  4.591 \\ 
KOI2711  & KOI0448  &    -40   &  5957 $\pm$ 110 &    4.44  $\pm$ 0.12 &    0.02  $\pm$ 0.09 &    1.07 $\pm$ 0.18 &  4.549 \\ 
KOI2711  & KOI0448  &    -30   &  5922 $\pm$ 110 &    4.46  $\pm$ 0.12 &   -0.01  $\pm$ 0.09 &    1.03 $\pm$ 0.18 &  4.466 \\ 
KOI2711  & KOI0448  &    -20   &  5894 $\pm$ 110 &    4.50  $\pm$ 0.12 &   -0.00  $\pm$ 0.09 &    1.00 $\pm$ 0.10 &  4.056 \\ 
KOI2711  & KOI0448  &    -10   &  5769 $\pm$ 110 &    4.51  $\pm$ 0.12 &   -0.01  $\pm$ 0.09 &    0.95 $\pm$ 0.10 &  2.899 \\ 
KOI2711  & KOI0448  &      0    & 5601 $\pm$ 110 &    4.49  $\pm$ 0.12 &   -0.11  $\pm$ 0.09 &    0.92 $\pm$ 0.10 &  1.907 \\ 
KOI2711  & KOI0448  &     10   & 5772 $\pm$ 110 &    4.51  $\pm$ 0.12 &    0.00  $\pm$ 0.09 &    0.95 $\pm$ 0.10 &  2.670 \\ 
KOI2711  & KOI0448  &     20   &  5878 $\pm$ 110 &    4.50  $\pm$ 0.12 &   -0.02  $\pm$ 0.09 &    0.99 $\pm$ 0.10 &  4.018 \\ 
KOI2711  & KOI0448  &     30   &  5939 $\pm$ 110 &    4.47  $\pm$ 0.12 &    0.02  $\pm$ 0.09 &    1.04 $\pm$ 0.18 &  4.511 \\ 
KOI2711  & KOI0448  &     40   &  5950 $\pm$ 110 &    4.44  $\pm$ 0.12 &    0.01  $\pm$ 0.09 &    1.06 $\pm$ 0.18 &  4.681 \\ 
KOI2711  & KOI0448  &     50   &  5953 $\pm$ 110 &    4.43  $\pm$ 0.12 &   -0.00  $\pm$ 0.09 &    1.07 $\pm$ 0.18 &  4.612 \\ 
KOI2711  & KOI0448  &     60   &  5948 $\pm$ 110 &    4.41  $\pm$ 0.12 &   -0.02  $\pm$ 0.09 &    1.09 $\pm$ 0.18 &  4.630 \\ 
KOI2711  & KOI0448  &     70   &  5922 $\pm$ 110 &    4.38  $\pm$ 0.12 &   -0.04  $\pm$ 0.09 &    1.13 $\pm$ 0.18 &  4.492 \\ 
KOI2711  & KOI0448  &     80   &  5916 $\pm$ 110 &    4.37  $\pm$ 0.12 &   -0.04  $\pm$ 0.09 &    1.13 $\pm$ 0.18 &  4.489 \\ 
KOI2711  & KOI0448  &     90   &  5943 $\pm$ 110 &    4.37  $\pm$ 0.12 &   -0.02  $\pm$ 0.09 &    1.13 $\pm$ 0.18 &  4.416 \\ 
KOI2711  & KOI0448  &    100   &  5946 $\pm$ 110 &    4.36  $\pm$ 0.12 &   -0.02  $\pm$ 0.09 &    1.14 $\pm$ 0.18 &  4.490 \\ 
KOI2711  & KOI0448  &    200   &  5977 $\pm$ 110 &    4.39  $\pm$ 0.12 &    0.02  $\pm$ 0.09 &    1.10 $\pm$ 0.18 &  4.478 \\ 
KOI2711  & KOI0448  &    300   &  5946 $\pm$ 110 &    4.37  $\pm$ 0.12 &   -0.01  $\pm$ 0.09 &    1.12 $\pm$ 0.18 &  4.476 \\ 
KOI2711  & KOI0448  &    400   &  5972 $\pm$ 110 &    4.36  $\pm$ 0.12 &   -0.01  $\pm$ 0.09 &    1.15 $\pm$ 0.18 &  4.412 \\
KOI2711  & KOI0448  &    500  &   5952 $\pm$ 110 &    4.38  $\pm$ 0.12 &   -0.02  $\pm$ 0.09 &    1.12 $\pm$ 0.18 &  4.361 \\ 
\enddata
\tablecomments{The first and second column list the Kepler Object of Interest (KOI) 
number of the primary and secondary star, respectively. The third column lists the 
RV shift that was applied to the scaled spectrum of the secondary star. 
The $\chi^2$ value is the median of the $\chi^2$ values returned by {\tt SpecMatch-Emp} for
each 100 {\AA} segment of the 5000$-$5800 {\AA} spectrum. \\
(This table is available in its entirety in machine-readable form.)}
\end{deluxetable*}

\newpage

\section{Results of Stellar Fits}
\label{results}

\begin{figure*}[!t]
\centering
\includegraphics[scale=0.63]{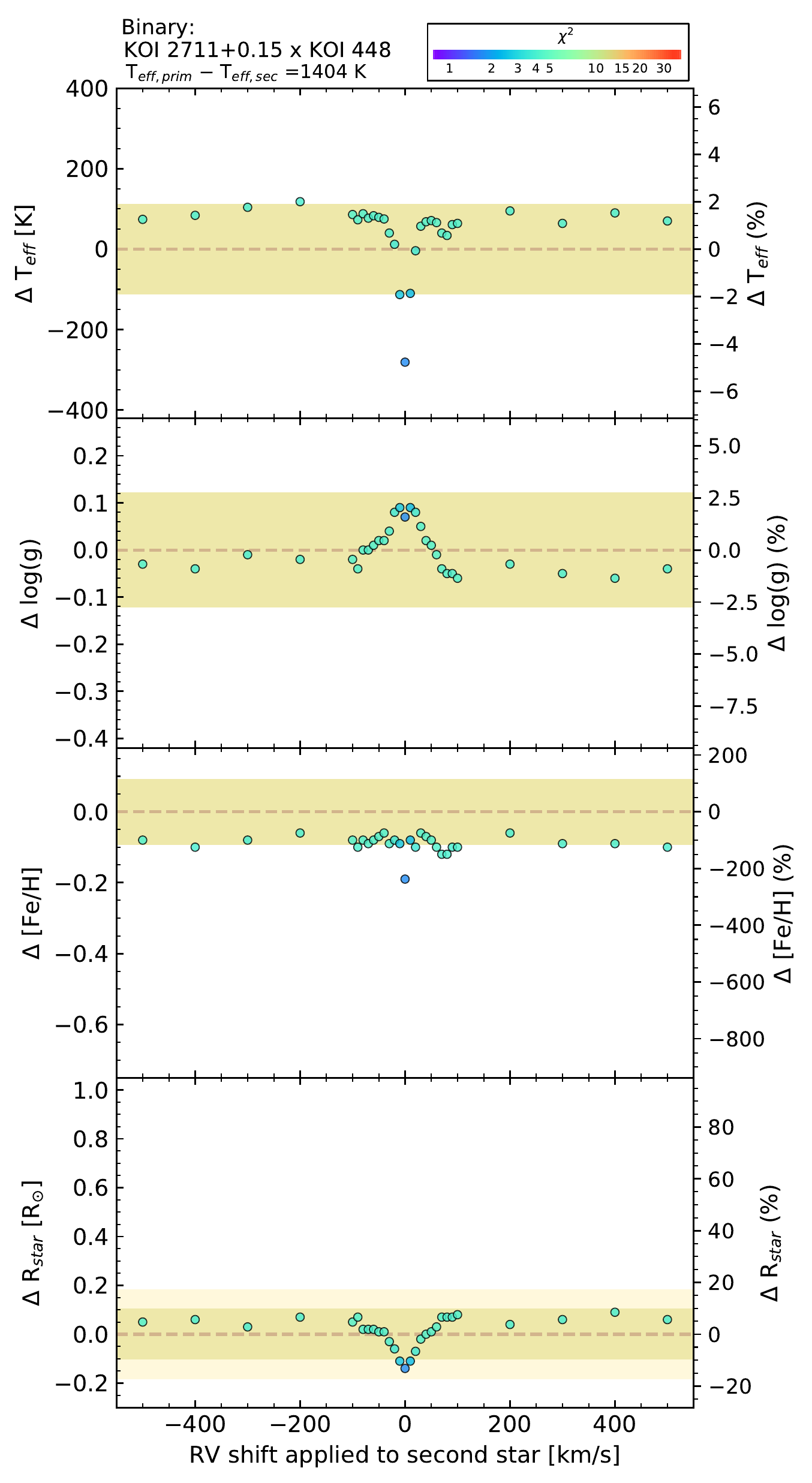}
\includegraphics[scale=0.63]{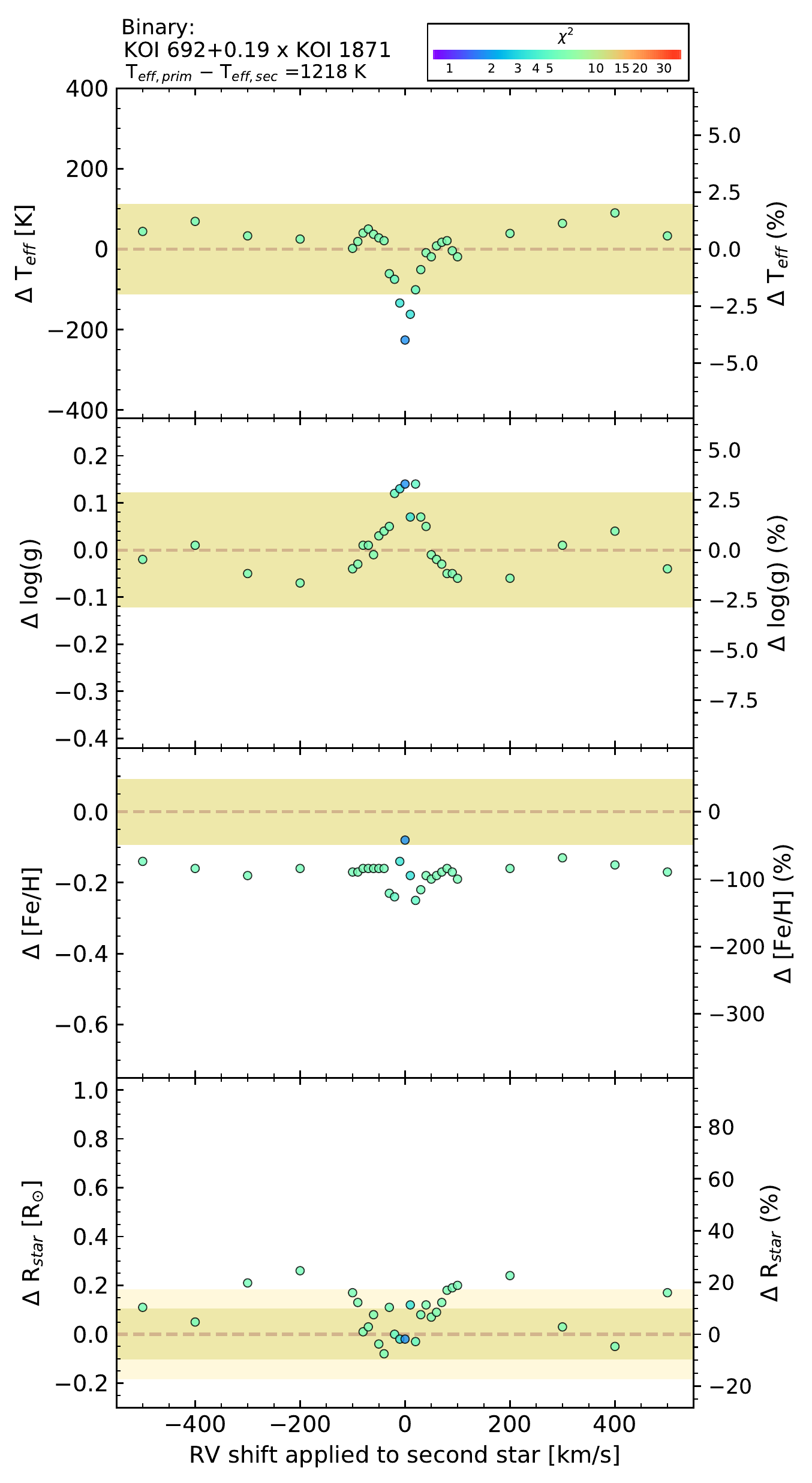}
\caption{Results from fitting a blended spectrum created by adding a scaled and RV-shifted
spectrum of a second (fainter) star to that of a primary star: differences in derived stellar 
parameters relative to the parameters of the primary star for Binary 1 (left) and Binary 2 (right)
as a function of the RV shift applied to the second star. The plotting symbols are color-coded
based on the $\chi^2$ value of the fit from {\tt SpecMatch-Emp} (see legend).
A positive value for a parameter difference means that the parameter derived from the 
blended spectrum is larger than the parameter derived from the spectrum of just the 
primary star. Conversely, a negative value means that the parameter derived from the 
unblended spectrum of the primary star is larger than the parameter derived from the 
blended spectrum.
The shaded area delineates the uncertainty associated with the various parameters as
returned by the {\tt SpecMatch-Emp} fits. The uncertainty for {\Rstar} can be either 0.1 or
0.18 {\Rsun}, shown as darker and lighter shaded areas, respectively.
\label{Fits_Binary_Panel1}}
\end{figure*}

\begin{figure*}[!t]
\centering
\includegraphics[scale=0.63]{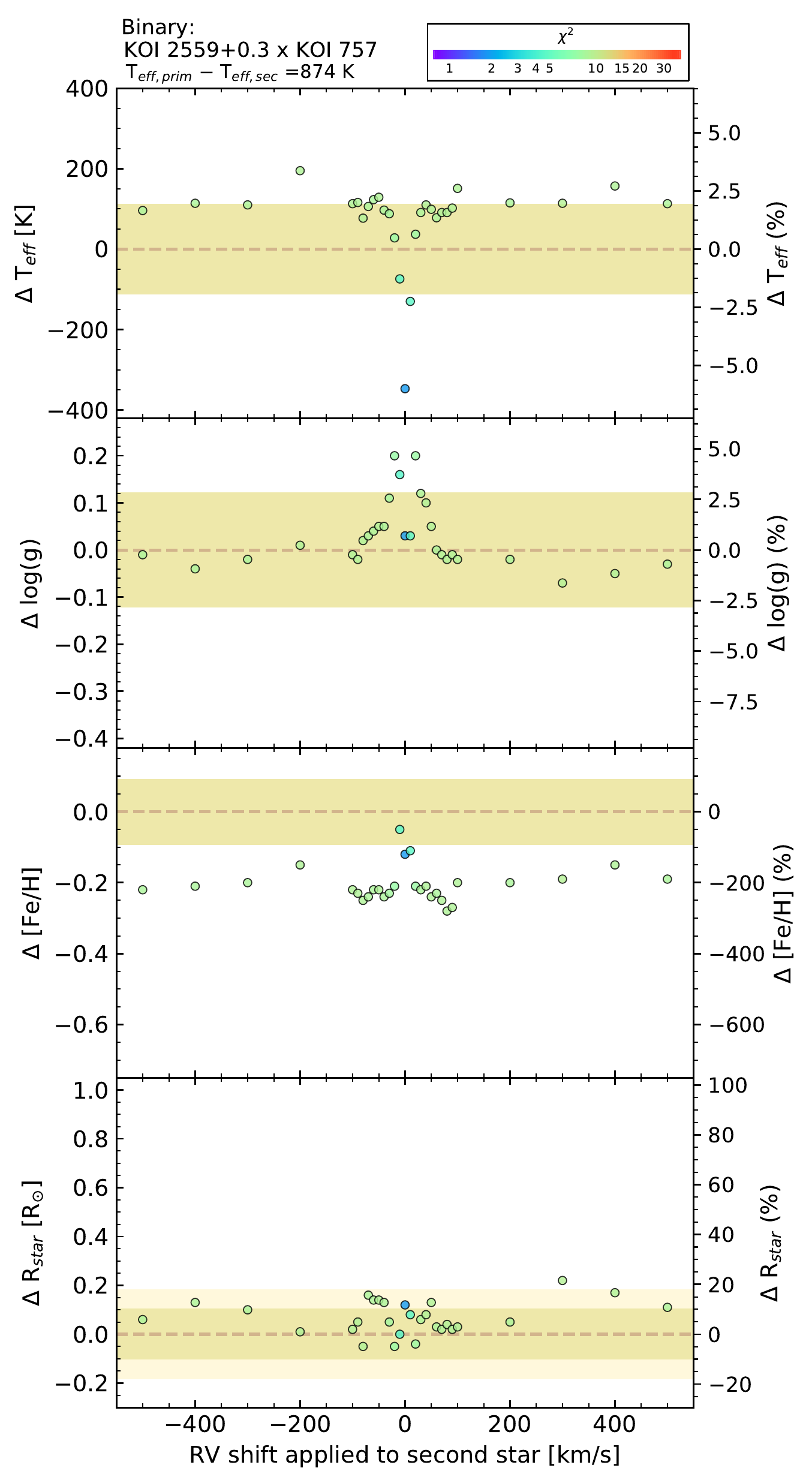}
\includegraphics[scale=0.63]{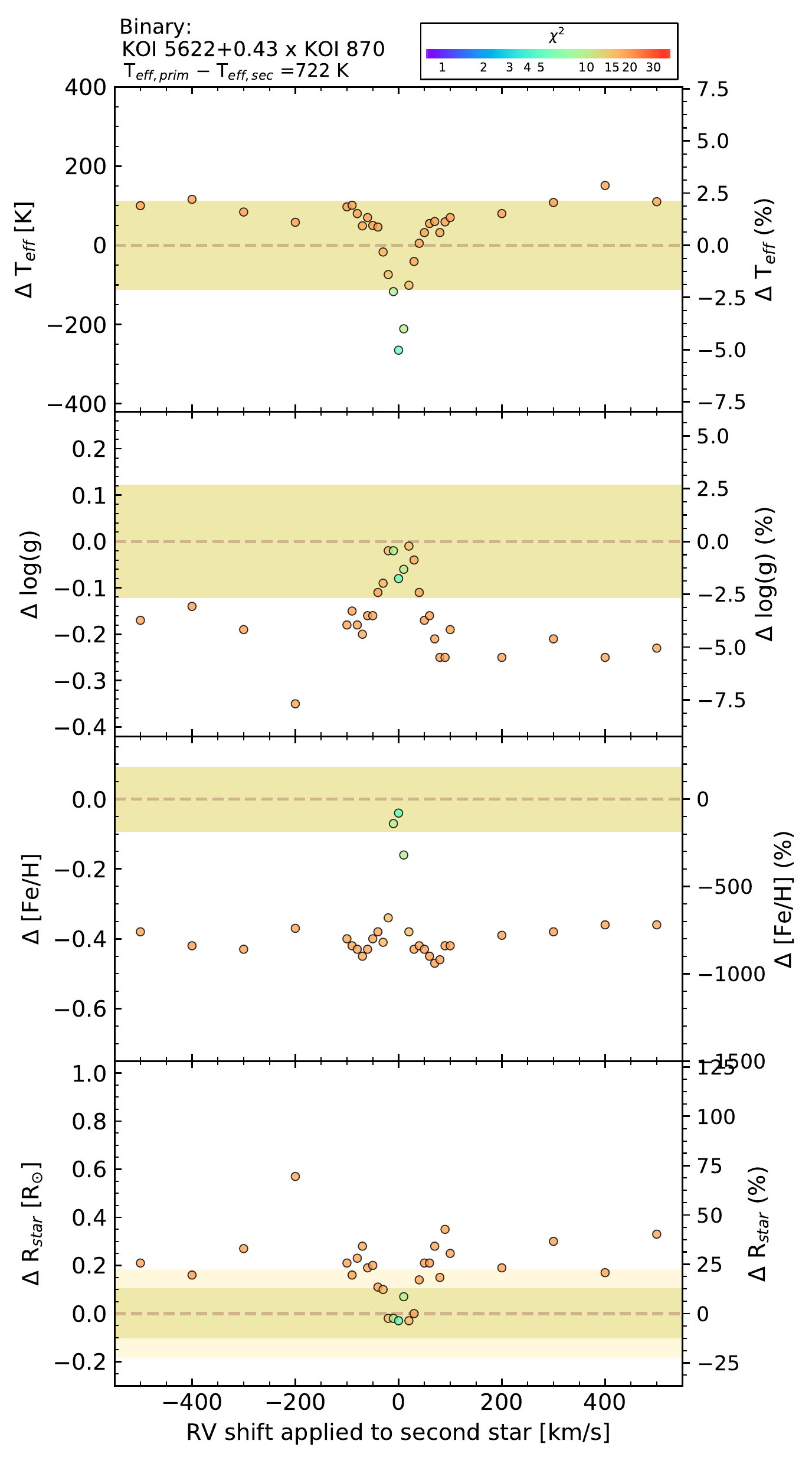}
\caption{Similar to Figure \ref{Fits_Binary_Panel1}, but for Binary 3 (left) and Binary 4 (right).
\label{Fits_Binary_Panel2}}
\end{figure*}

\begin{figure*}[!t]
\centering
\includegraphics[scale=0.63]{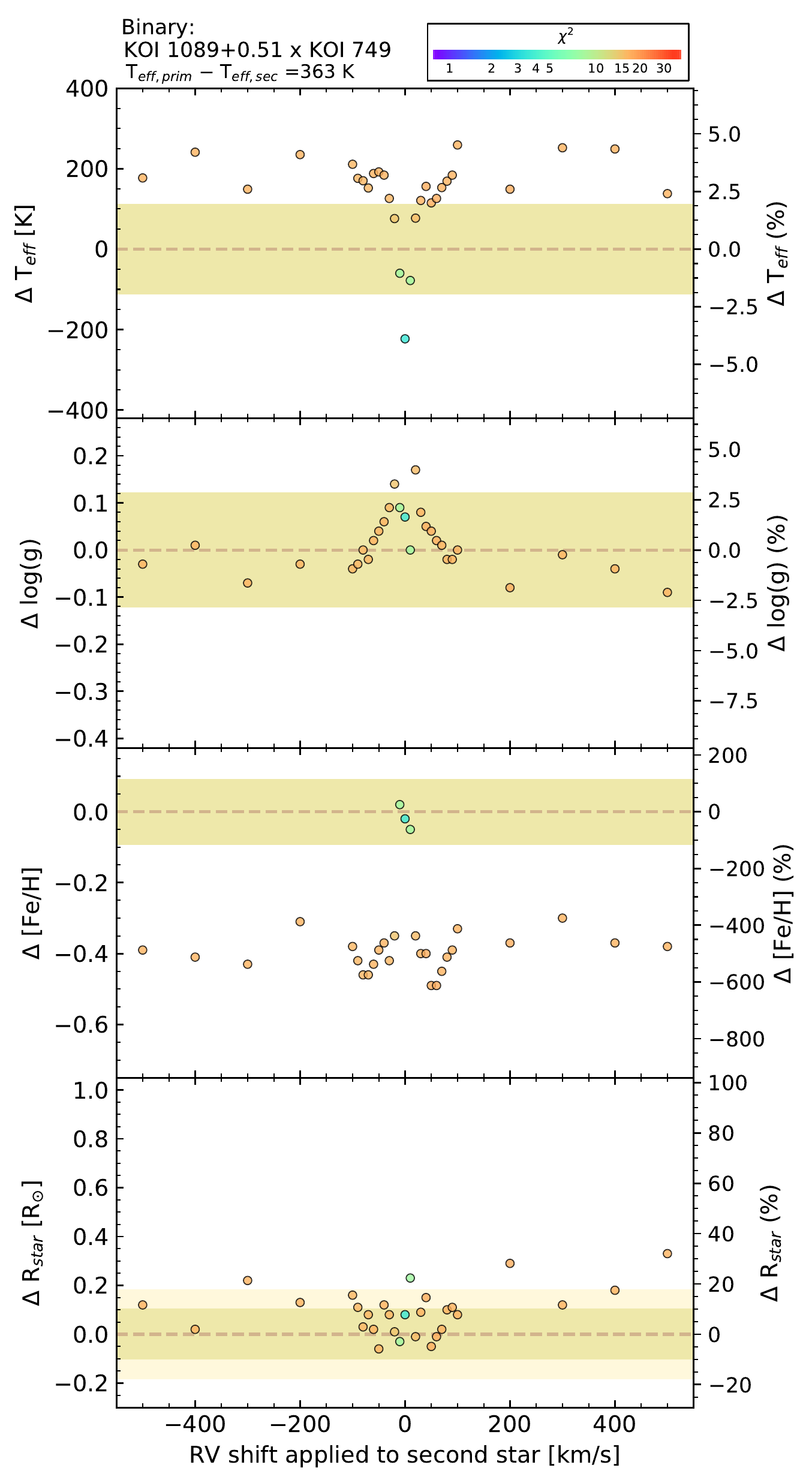}
\includegraphics[scale=0.63]{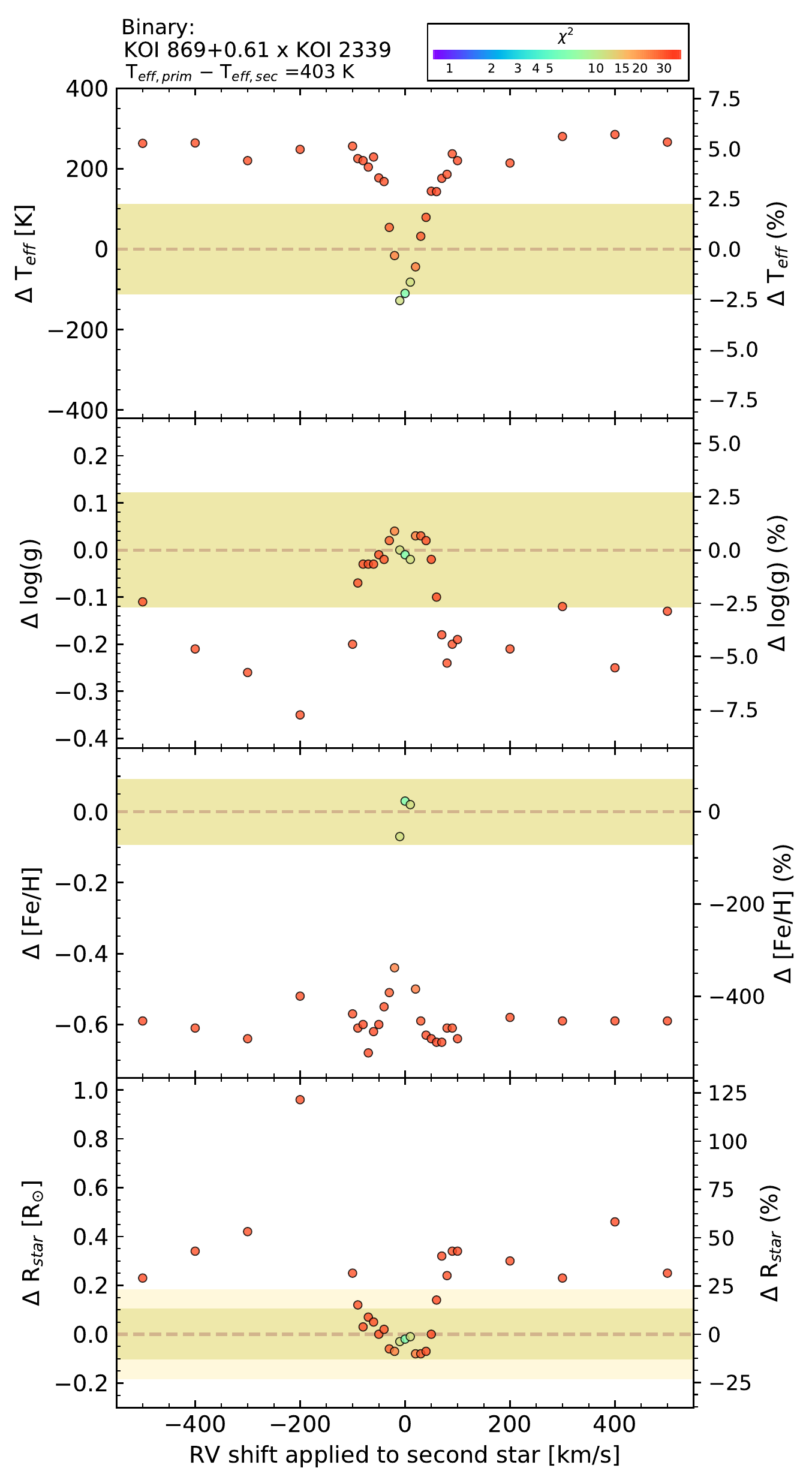}
\caption{Similar to Figure \ref{Fits_Binary_Panel1}, but for Binary 5 (left) and Binary 6 (right).
\label{Fits_Binary_Panel3}}
\end{figure*}

\begin{figure*}[!t]
\centering
\includegraphics[scale=0.63]{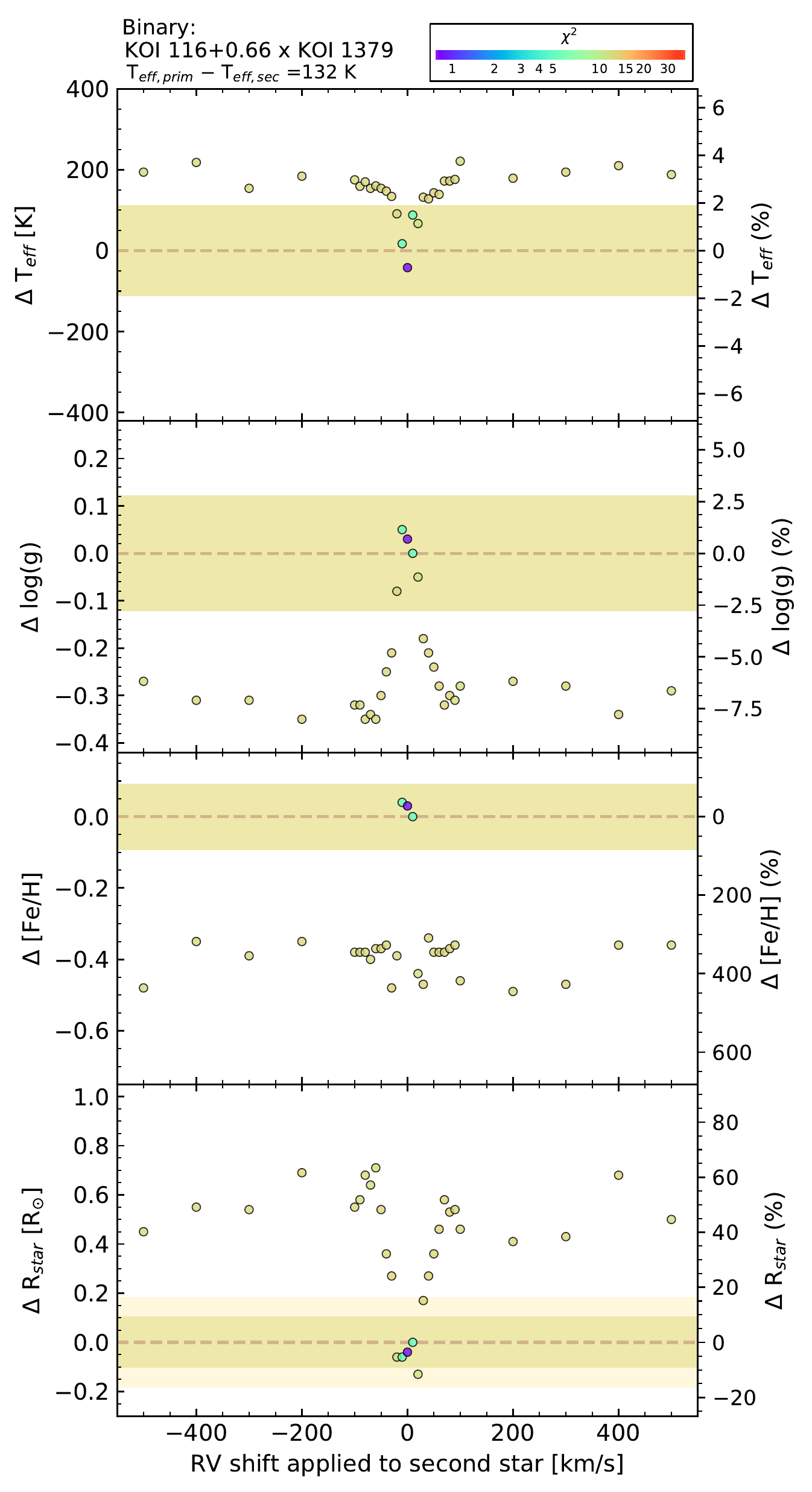}
\includegraphics[scale=0.63]{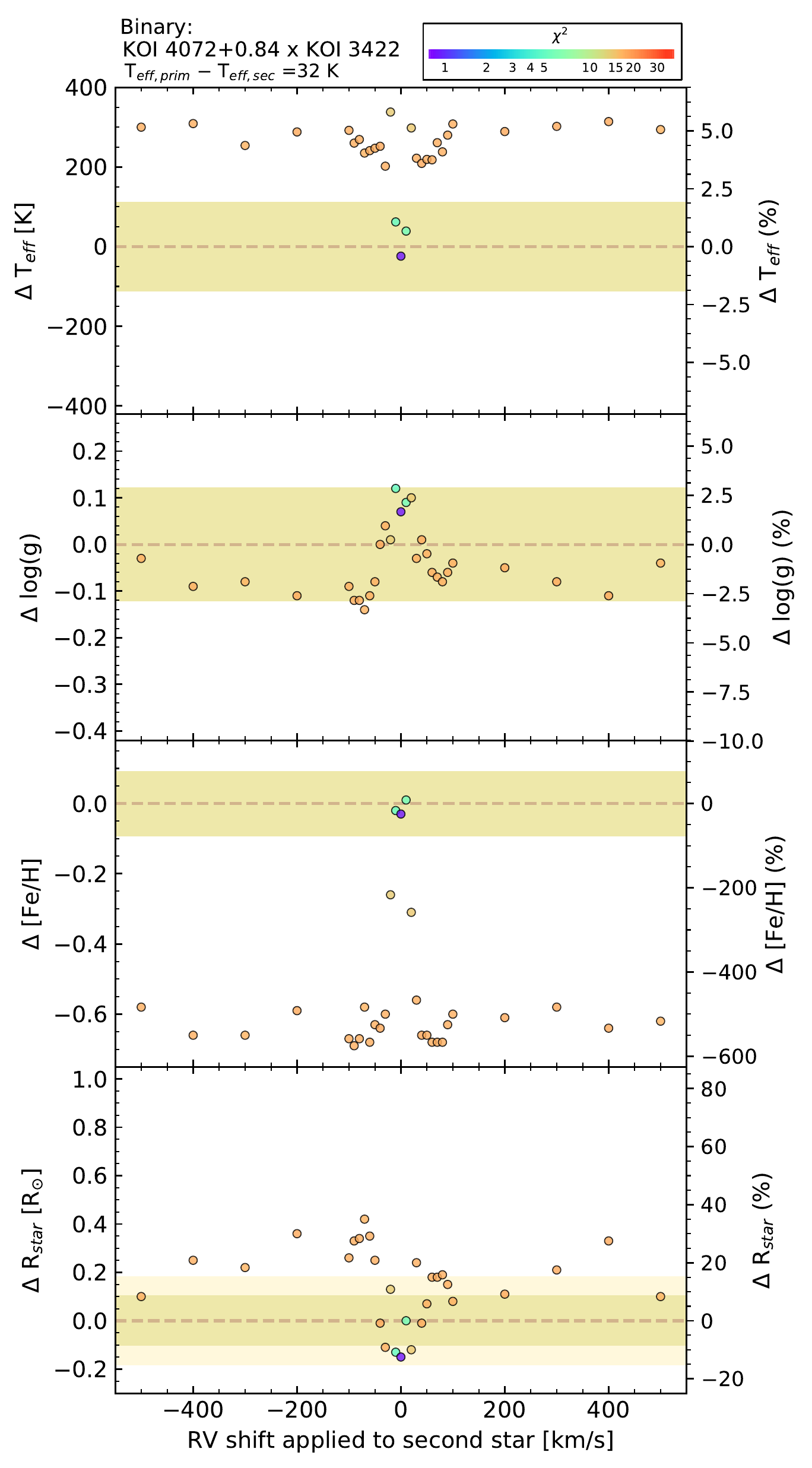}
\caption{Similar to Figure \ref{Fits_Binary_Panel1}, but for Binary 7 (left) and Binary 8 (right).
\label{Fits_Binary_Panel4}}
\end{figure*}

To quantify the effect of deriving stellar parameters from blended spectra, we used
the 8 pairs of spectra and created 29 simulated binaries for each pair: the spectrum
of the cooler star was multiplied by the luminosity ratio of the two stars and also shifted 
in wavelength space by 29 different RV values (see section \ref{sample}) before 
co-adding it to the spectrum of the 
hotter star. After creating these blended spectra, we fit them with {\tt SpecMatch-Emp} 
to derive $T_{\mathrm{eff}}$, $\log (g)$, [Fe/H], and the stellar radius ({\Rstar}). 
The results of the stellar fits for each simulated binary are listed in Table
\ref{star_params_binaries} and shown in Figures \ref{Fits_Binary_Panel1} to 
\ref{Fits_Binary_Panel4}, where the difference between the stellar parameters 
derived from the blended spectrum and the parameters of the primary star are 
plotted as a function of RV shift applied to the secondary star.

The data points in these figures are color-coded based on the unnormalized
$\chi^2$ value of the fit, which was calculated based on the output of {\tt SpecMatch-Emp}. 
Since for the fits the 5000-5800 {\AA} spectra are divided into 100 {\AA} segments, 
{\tt SpecMatch-Emp} returns eight $\chi^2$ values for each fit; they are usually very 
similar, but to prevent outliers from inflating a $\chi^2$ value, we calculated the 
median of the eight $\chi^2$ values. Even when individual stars are fit, the $\chi^2$ 
value is not necessarily close to 0 (see Table \ref{star_params}); values 
larger than $\sim$ 15 likely suggest a bad fit. Thus, stellar parameters from fits 
with large $\chi^2$ values are in general less reliable. It appears that for most of 
the simulated binaries with a difference between effective temperatures of the 
two stars $\lesssim$ 750 K and RV shifts of the secondary star $\gtrsim$ 
10 km s$^{-1}$, the fits have large $\chi^2$ values.

For all simulated binaries, the discrepancies in derived stellar parameters are largest
for the metallicities: the [Fe/H] values tend to be smaller when derived from a blended 
spectrum by factors of a few relative to the values derived for the primary star alone, 
with the effect being larger the smaller the difference in luminosity between the primary 
and secondary star. Also, introducing RV shifts larger than 10 km s$^{-1}$ to the 
spectrum of the second star results in a substantial decrease in derived [Fe/H] values, 
but this trend plateaus beyond about 100 km s$^{-1}$. Only no RV shift or RV shifts 
up to 10 km s$^{-1}$ will result in [Fe/H] values consistent with values derived from
an unblended spectrum of the primary star. However, for the brighter companion stars, 
the derived stellar parameters are also more unreliable, as gauged by the $\chi^2$ 
value returned by the fit.

\newpage

The surface gravities and effective temperatures tend to agree within $\sim$ 5\% with 
the values derived for the unblended primary star. Deviations in $T_{\mathrm{eff}}$ are 
larger when no or a very small RV shift is included for the companion, but only for the 
simulated binaries whose components have effective temperatures that differ by $\gtrsim$ 
700 K (Binaries 1-4; see Table \ref{star_params}). In these cases, the effective 
temperature is underestimated for the blended spectrum by up to a few hundred K; 
it is more in agreement with the value determined from the unblended spectrum 
when the companion spectrum is shifted by $>$ 20 km s$^{-1}$. The $\log (g)$ values 
do not show significant deviations for most binaries (the uncertainty returned by 
{\tt SpecMatch-Emp} is 0.12); only some blended spectra cause $\log (g)$ values 
to be underestimated, but these values are also more unreliable.

The stellar radii derived from blended spectra agree quite well with those derived
from the primary spectrum alone (considering typical uncertainties of 0.18 \Rsun\ 
from the fits), except for certain binaries when RV shifts are introduced. In such 
cases, the radii end up being overestimated, with typically larger values for larger
RV shifts (up to $\sim$ 100 km s$^{-1}$), but, as with the $\log (g)$ values, they 
are less reliable. At smaller RV shifts, stellar radii tend to be somewhat underestimated,
but by less than the uncertainty of 0.18 typically returned by {\tt SpecMatch-Emp}.

In general, when the temperature difference between the primary and the companion 
star is $\lesssim$ 750 K (Binaries 4-8), stellar parameters derived from blended spectra
show more significant deviations when RV shifts greater than a few tens km s$^{-1}$ 
are introduced. Effective temperatures are overestimated by up to $\sim$ 300 K, $\log (g)$
values underestimated by up to $\sim$ 0.2-0.3 dex, [Fe/H] values are underestimated by
0.4-0.7 dex, and {\Rstar} is overestimated by up to $\sim$ 60\%.
 
\begin{figure*}[!t]
\centering
\includegraphics[scale=0.7]{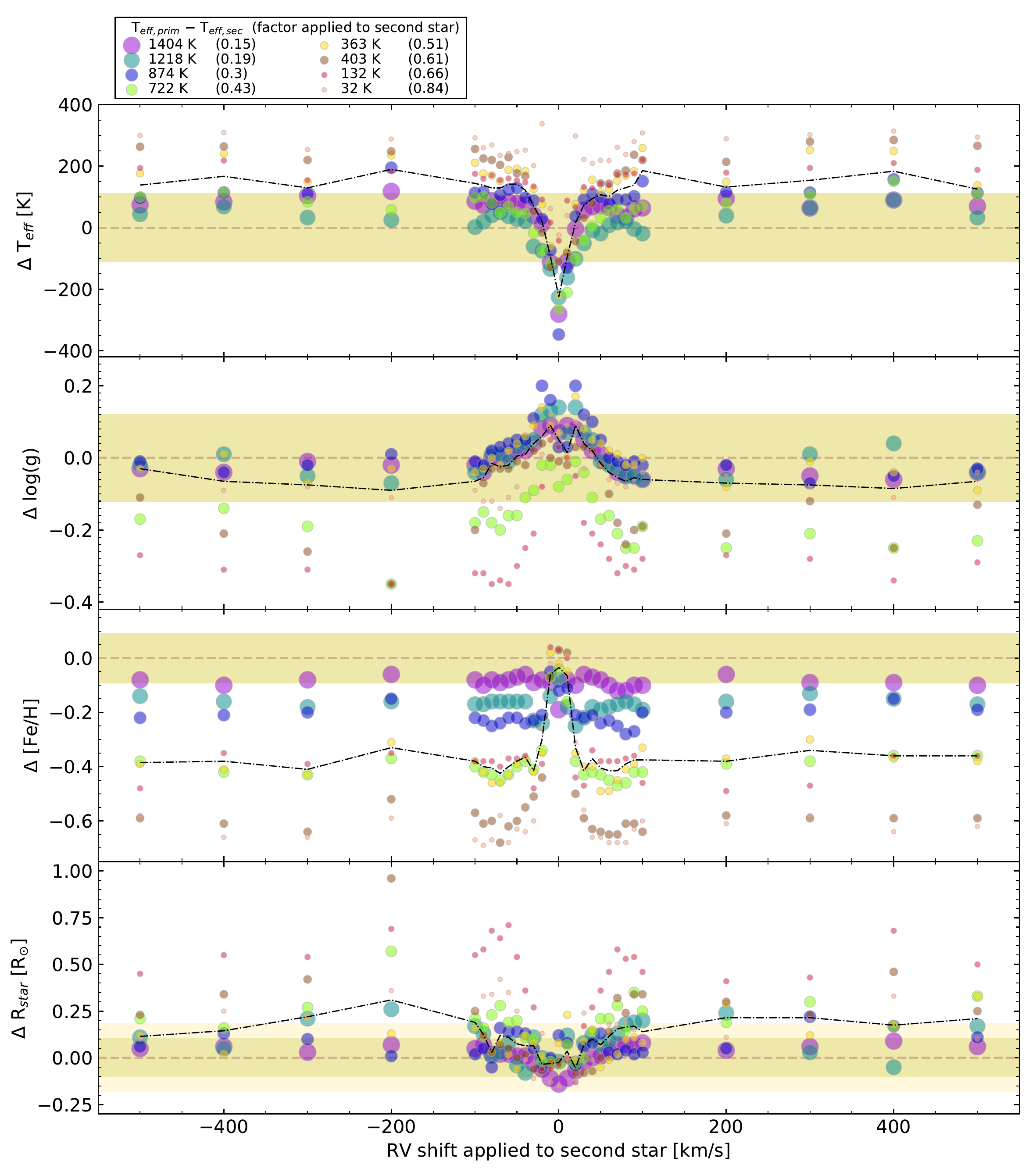}
\caption{Results from Figures \ref{Fits_Binary_Panel1} to \ref{Fits_Binary_Panel4} combined 
in one plot. The colors and sizes of the plotting symbols vary with the difference in $T_{\mathrm{eff}}$ 
between the primary and secondary star: the larger symbols represent binaries with larger
temperature differences between the two stars. The median differences in stellar parameters 
as a function of RV shift are shown with the black dash-dotted line. The shaded areas have the
same meaning as in Figures \ref{Fits_Binary_Panel1} to \ref{Fits_Binary_Panel4} (see
caption of Fig.\ \ref{Fits_Binary_Panel1}).
\label{Fit_results_all}}
\end{figure*}

Figure \ref{Fit_results_all} combines the results shown in Figures \ref{Fits_Binary_Panel1} 
to \ref{Fits_Binary_Panel4}; the median differences in parameter values are shown as
a dash-dotted line.
The large deviations in $T_{\mathrm{eff}}$ for simulated binaries with no RV shift for the 
companion star are apparent, as well as the large underestimated [Fe/H] values, which
tend to be larger the brighter the companion star is (i.e., the smaller the difference in 
effective temperatures between the two stars). Metallicities, surface gravities, and 
stellar radii are least affected if no RV shift is present.
Considering the uncertainties in the stellar parameters returned by the fit 
($\sigma(T_{\mathrm{eff}})=$ 110 K, $\sigma(\log (g))=$ 0.12, $\sigma$([Fe/H]) 
$=$ 0.09, $\sigma(R_{\star})=$ 0.1-0.18 {\Rsun}), the $\log (g)$ values are overall least affected 
by adding a second stellar spectrum to that of the primary star. Median differences in
parameter values for $\log (g)$ are within 1$\sigma$; those for $T_{\mathrm{eff}}$ 
and {\Rstar} are typically just larger than 1$\sigma$, while those for [Fe/H] are at 
about 4$\sigma$. However, especially for binaries with brighter companion stars, 
[Fe/H] can be underestimated by up to 8$\sigma$ and {\Rstar} overestimated by
up to 4$\sigma$. 

From Figure \ref{Fit_results_all}, it is clear that stellar radii are significantly different only
when the spectrum of a companion star is added with an RV shift of $\gtrsim$ 50 km s$^{-1}$
and also a difference in $T_{\mathrm{eff}}$ with respect to the primary of 720 K or $\lesssim$ 
400 K. For smaller or no RV shifts, and for simulated binaries with fainter secondaries, the 
stellar radius derived from the blended spectrum is not significantly different from the radius 
derived from the uncontaminated spectrum. 
However, the radius uncertainties as determined by {\tt SpecMatch-Emp} are typically
0.18 {\Rsun} for the stars we selected as primaries (for stellar radii ranging from 0.8 to 
1.2 {\Rsun}, with a median value of 1.06 {\Rsun}; see Table \ref{star_params}), so of the
order of 17\%. These radii are just derived by comparing the observed spectrum to the
library spectra and using a weighted average of the stellar radii of the five best-matching 
library stars. So, the radius uncertainties are set by the accuracy of the library parameters,
and are relatively large. 
Other methods, like using isochrone fitting to convert $T_{\mathrm{eff}}$, $\log (g)$, and 
[Fe/H] to {M$_{\ast}$}, {\Rstar}, and age, can yield more precise stellar radii, with uncertainties
of $\sim$ 11\% \citep{johnson17} or even 3\% (when {\it Gaia} parallaxes are used;
\citealt{fulton18}). Using these methods, the effect of a blended companion star on the 
derived stellar radius might be more noticeable.

\clearpage

\section{Discussion}
\label{discuss}

\begin{figure*}[!th]
\centering
\includegraphics[scale=0.5]{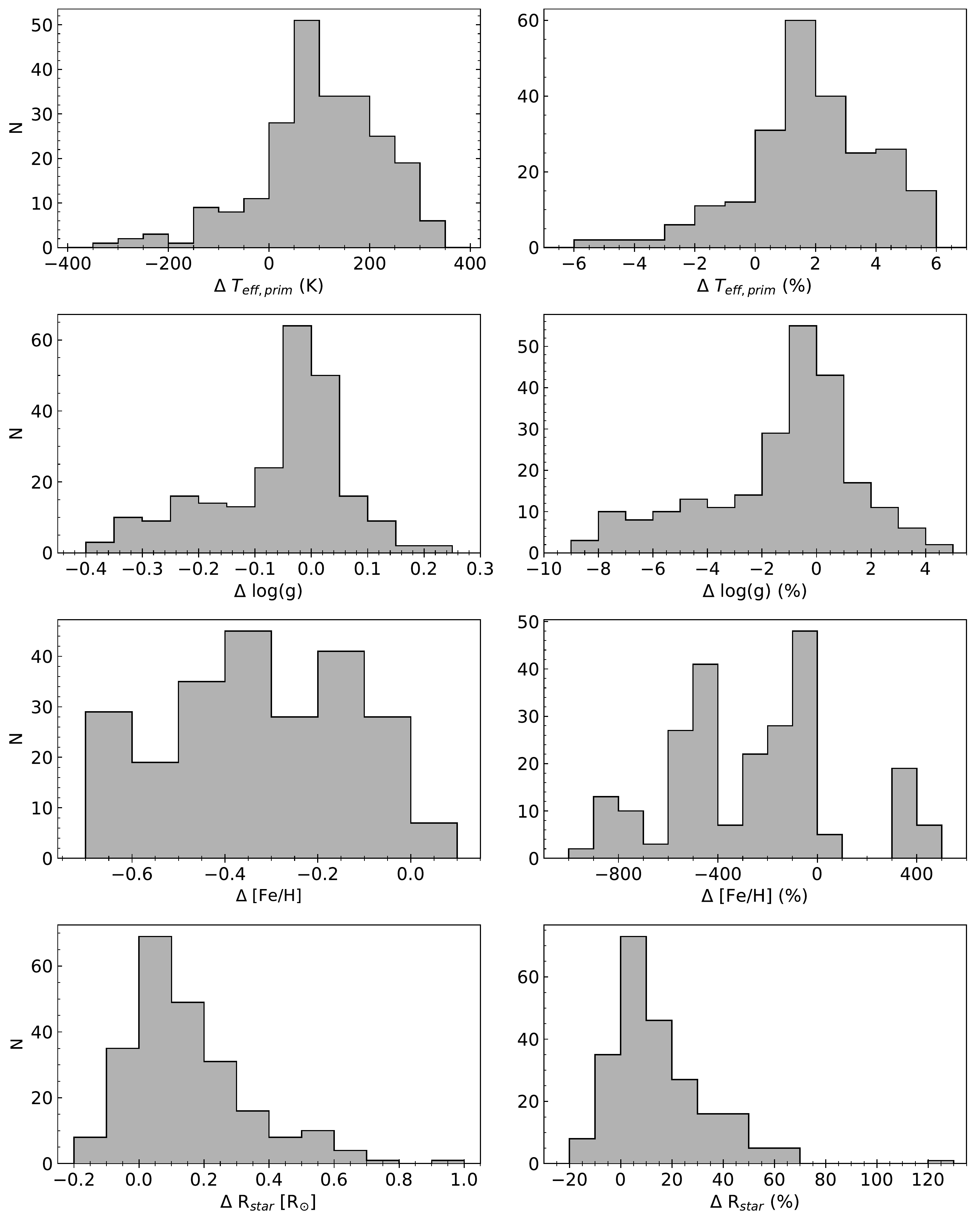}
\caption{Histograms of the stellar parameter differences for the primary star (calculated 
as stellar parameters derived from a blended spectrum minus stellar parameters derived 
from the single spectrum; same data points as in Figures 
\ref{Fits_Binary_Panel1}--\ref{Fits_Binary_Panel4}) for all 232 simulated binaries. 
The panels on the left show the differences in values, while the panels on the right 
show the percentage differences (with respect to the value derived from the single spectrum 
of the primary star). In all panels, a value of zero means that the derived parameter is 
accurate for the primary star.
\label{Fit_results_histo}}
\end{figure*}

\begin{figure*}[!t]
\centering
\includegraphics[scale=0.47]{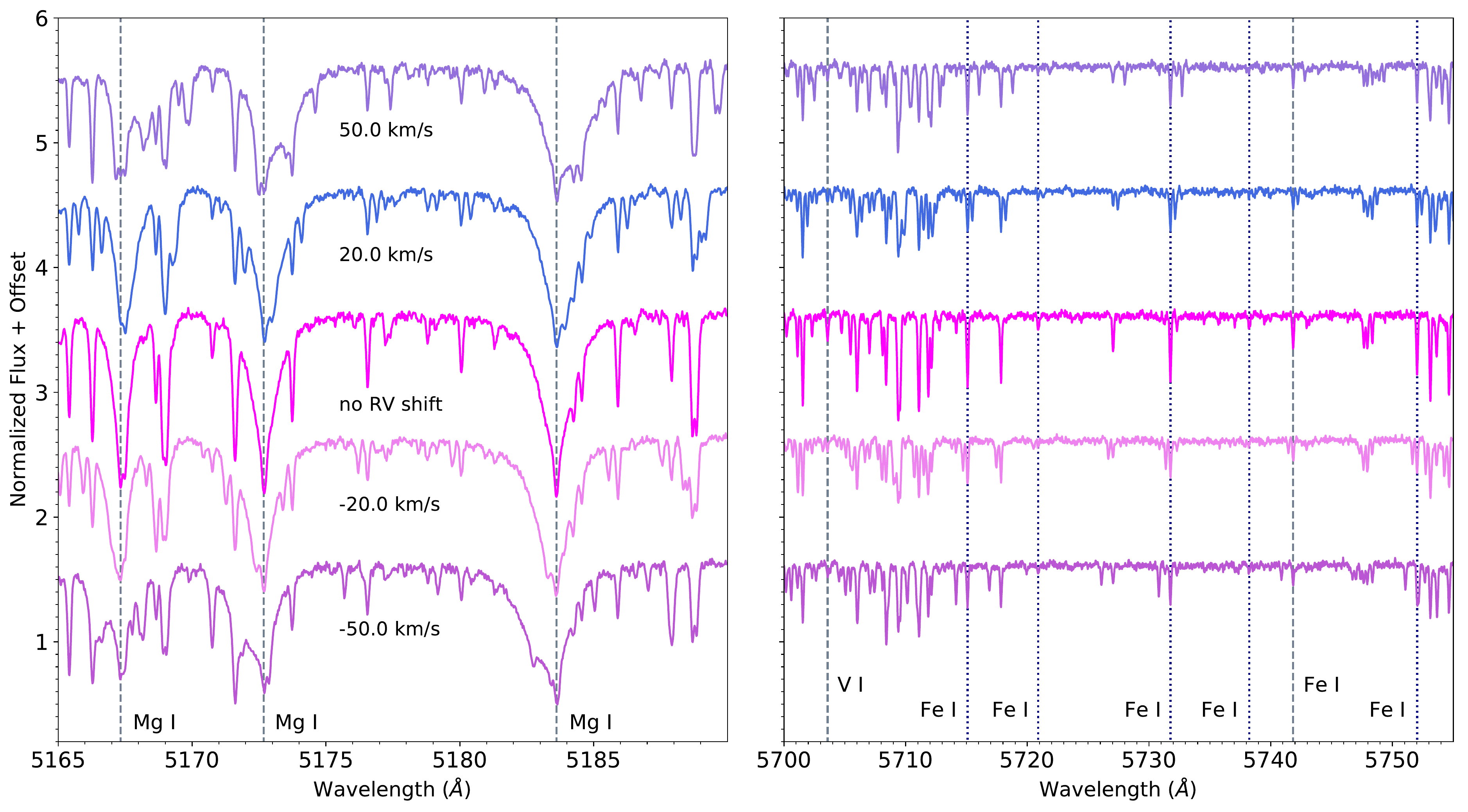}
\caption{Two spectral regions of a simulated binary (Binary 7), where the spectrum of the 
cooler star has been multiplied by a factor equal to the luminosity ratio of the two stars and
shifted by different RV values as indicated in the figure label. A few metal lines that serve
as indicators for surface gravity (Mg I), metallicity (Fe I), or effective temperature (e.g.,
ratio of V I and Fe I line at 5703.6 and 5741.9 {\AA}, respectively) are also labeled.
\label{binary_lines_example}}
\end{figure*}

Fitting a spectrum of a stellar blend with {\tt SpecMatch-Emp} assuming only 
one star is present will result in stellar parameters that may be inaccurate, deviating 
by more than the 1-$\sigma$ uncertainties returned by the fit (see Figure 
\ref{Fit_results_histo}). If the companion star is faint, the effect on the derived 
stellar parameters of the primary star is very minor. At a minimum, the contaminating 
star will add some excess noise to the spectrum of the primary star. A brighter 
companion affects the determination of stellar parameters in non-intuitive ways -- 
its effects depend on the properties of the two stars and how their spectral signatures 
are blended.
The $\chi^2$ value returned by the fit can be used as an indicator of whether stellar 
parameters are still reliable, even though the largest $\chi^2$ values are not always
associated with the largest discrepancies in derived stellar parameters.

From our analysis of 232 simulated binaries, we find that, if the secondary star's spectrum 
is shifted in wavelength due to a radial velocity difference with respect to the primary star, 
in some cases the stellar parameters of the primary star can still be retrieved reliably. 
It all depends whether {\tt SpecMatch-Emp} can still determine a similar set of 
best matching spectra as for the unblended spectrum of the primary star. While there 
are several line diagnostics that are sensitive to the various stellar parameters, such 
as ratios of weak metal lines (often of the same element and nearby in wavelength space) 
for $T_{\mathrm{eff}}$, the Mg I b triplet for $\log (g)$, and various iron lines for [Fe/H]
\citep[e.g.,][]{gray94,gray96,sousa10}, it is difficult to determine which set of spectral 
lines most influences the outcome of {\tt SpecMatch-Emp}. A few examples of how 
adding an RV-shifted spectrum of a second star to that of the primary star affects some 
of the spectral lines are shown in Figure \ref{binary_lines_example}. The line profile of the
Mg I b triplet clearly changes, and the depth of other metal lines is typically reduced.
In general, weaker Mg I and Fe I lines would imply lower surface gravities and 
metallicities, respectively. It is also worth noting that altered line profiles, which would
vary over time as the two stars orbit their center of mass, could affect the precision 
of RV measurements needed to measure exoplanet masses.

\begin{figure*}[!t]
\centering
\includegraphics[scale=0.6]{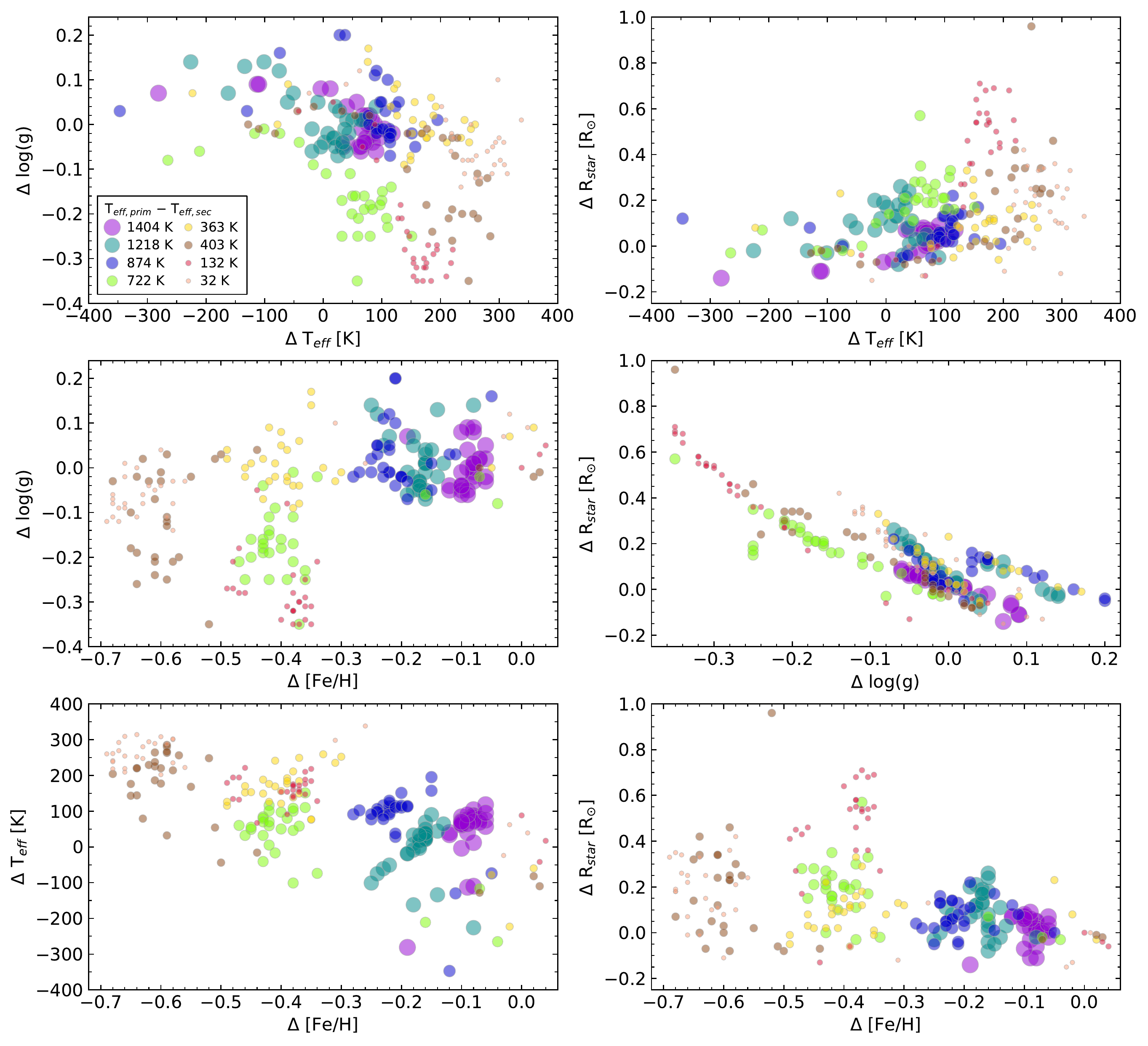}
\caption{Comparison of the $T_{\mathrm{eff}}$, $\log (g)$, [Fe/H], and {\Rstar} 
differences for the primary star (parameters derived from a blended spectrum 
minus the parameters derived from the single spectrum, as in Figures 
\ref{Fits_Binary_Panel1} to \ref{Fits_Binary_Panel4}), with symbol sizes and 
colors according to the difference in $T_{\mathrm{eff}}$ between the primary 
and secondary star (larger symbols for larger temperature differences). 
\label{Fit_results_comp}}
\end{figure*}

To see any correlations in how the various stellar parameters change due to 
a blended spectrum, in Figure \ref{Fit_results_comp} we compare the differences 
between $T_{\mathrm{eff}}$, $\log (g)$, [Fe/H], and {\Rstar} when these parameters 
are derived from a blended spectrum and when derived from the original spectrum of
the primary star. When a blended spectrum results in lower $\log (g)$ values, the 
derived [Fe/H] values are also smaller, but the derived $T_{\mathrm{eff}}$ values 
are typically larger. This mainly occurs when the luminosity ratio between secondary 
and primary star is $\gtrsim$ 0.4 (or the temperature difference between primary and
secondary star is $\lesssim$ 700 K). When $T_{\mathrm{eff}}$ values are underestimated, 
[Fe/H] values are still underestimated, but by smaller amounts, and $\log (g)$ values 
are roughly in agreement with the values derived from the unblended spectrum. 
The differences in stellar radii are correlated with the differences in $\log (g)$,
as expected ($\log (g) \propto R_{\ast}^{-2}$); trends for the metallicity and effective
temperature are less clear, except that the largest deviations in stellar radius occur
when $T_{\mathrm{eff}}$ is overestimated by $\sim$ 150-250 K.
This demonstrates that adding the spectrum of a second star above a certain 
brightness level will reduce the strength of lines of both pressure-sensitive and 
metallicity-sensitive lines, making stars appear to have lower surface gravity (or a 
larger radius) and lower metallicity. In these cases the star's effective temperature is 
typically overestimated.

The primary star's effective temperature still agrees with the value derived from an 
unblended spectrum when the companion star is less than half as bright as the 
primary and its spectrum shifted by more than $\sim$ 20 km s$^{-1}$. This is likely 
a result of temperature-sensitive lines of the primary star not being blended or 
distorted by the sufficiently shifted lines of the secondary star. If the two stars are 
very similar, only an RV shift of 0 results in still accurate parameters for the primary 
star (see Figure \ref{Fit_results_all}). In this case the spectral features of the 
primary and secondary star overlap, and the fitting routine will still find an accurate 
match for the primary star. 
A blended companion star will cause the effective temperature of the primary star to not 
deviate more than 6\%, which is relatively small, but triple the $T_{\mathrm{eff}}$ 
uncertainty returned by {\tt SpecMatch-Emp} (see also Figure \ref{Fit_results_histo}). 
This result is encouraging for studies that rely on temperature estimates from spectroscopy,
such as the derivation of accurate $\log (g)$ and {\Rstar} values from asteroseismology
\citep[e.g.][]{huber13}. The presence of a companion star will make the effective 
temperature more unreliable, but for most cases it is expected to lie within 2-3 $\sigma$
of the actual value.
 
Surface gravities and stellar radii seem to be least affected when a blended spectrum
is fit, especially when considering the $\sim$ 3\% and $\sim$ 15\%, respectively, 
uncertainty in these parameters returned by {\tt SpecMatch-Emp}. Fitting a blended
spectrum results in deviations of $\log (g)$ and {\Rstar} values of at most $\sim$ 2 
$\sigma$ in almost all cases (see also Figure \ref{Fit_results_histo}). 
However, as mentioned earlier, different methods to derive {\Rstar}, like using isochrone 
fitting, yield more precise stellar radii, but they also rely on other spectroscopically derived 
parameters. In this case the uncertainties in {\Rstar} would be dominated by the uncertainties 
in these parameters, in particular the effective temperature \citep[see][]{fulton18}. Thus, 
even if a precision of 3\% could be achieved for {\Rstar}, a deviant $T_{\mathrm{eff}}$ 
value due to a blended companion spectrum could increase the uncertainty of the 
stellar radius by several percentage points.

The stellar radius is an important parameter, especially for transiting planets,
where the planet radius scales with the stellar radius. Any increase in the uncertainty
of the stellar radius raises the uncertainty of the planet radius, thus affecting our
interpretation of its properties, such as its density and atmospheric scale height,
as well as the distribution of planet sizes \citep[e.g.,][]{weiss14,rogers15,fulton17}.
Radius uncertainties of $\sim$ 10\% or less are needed to see features and trends
in planet radii distributions \citep{fulton17,fulton18}.
Planets transiting a star with an overestimated radius have derived planet radii that
are too large, too (since $R_{\mathrm{planet}} \propto R_{\star}$), and thus may
be interpreted as having more volatiles than is actually the case. The stellar radius 
tends to be overestimated if a similar, bright star with RV shifts larger than a few
tens of km s$^{-1}$ is blended with the primary star. Such a bright, nearby companion 
could be detected in high-resolution images or identified by its set of absorption lines 
in the blended spectrum; faint companions are more difficult to detect spectrally
(becoming essentially undetectable once their luminosity ratio with respect
to the primary drops below $\sim$ 10\%).

Metallicities are generally unreliable when derived from a stellar blend, even if the two
stars have very similar metallicities. Unless the companion star is faint ($\sim$ 0.1 the
luminosity of the primary) or has no or a very small RV shift relative to the primary, 
the metallicity derived from the combined spectrum is smaller by up to 7 times the typical 
uncertainty in [Fe/H] values for most cases. Indeed, once the RV shift is larger than
about 50 km s$^{-1}$ (which corresponds to shifts in wavelength of $\gtrsim$ 1 {\AA}), the 
metallicity is underestimated by a constant amount; this results from the Fe I lines being
diluted (and not distorted) by the light of the companion.
Underestimating stellar metallicities could skew planet population studies, which have
shown that the occurrence of large planets (from sub-Neptunes to Jupiters) increases
with stellar metallicity \citep{santos04,fischer05,johnson10,wang15,petigura18,narang18}, 
while small, rocky planets ($R_p \lesssim$ 1.7 {\RE) are found around stars with a wider 
range of metallicities \citep{buchhave12,everett13,wang15,petigura18}. 
Higher actual metallicities for some stars could weaken the trend seen for some 
super-Earths and sub-Neptunes. 
In addition, the metallicity of a star is used to inform planet formation models, where
the metallicity of the protoplanetary disk is assumed to be the same as that of the star,
and so more metal-rich stars are assumed to have disks with a higher solid surface
density. In general, rocky cores of gas giant planets are thought to form more efficiently 
around metal-rich stars, thus allowing substantial atmospheres to be accreted 
\citep[e.g.,][]{dawson15}. Some models explaining the formation of super-Earths and 
sub-Neptunes might have to be adjusted if the metallicity of their host stars were
underestimated.
Finally, if certain stars are wrongly determined to be metal-poor, they may be seen as 
unlikely hosts for giant planets, so they could potentially be left out of target lists for 
planet searches or follow-up observations. 

In summary, we note that the deviations introduced in the stellar parameters from 
fitting a blended spectrum strongly depend on the characteristics of this spectrum; 
the absorption lines from both stars, in particular whether the lines from the primary star 
are still identifiable by the fitting routine, as well as their signal-to-noise ratio, will influence 
the outcome of the stellar fit. Based on the results of our 232 simulated binaries, we 
observed some trends, but there are also features that really depend on the two individual 
spectra that were combined. Thus, it is difficult to accurately predict for any specific case 
by how much the presence of a contaminating star will cause a deviation in the derived 
stellar parameters; in general, stellar parameters become more unreliable.
In a bound binary system, the relative radial velocity between the two stars 
will rarely be 0 due to their orbital motion; when we observe such a system, it is more 
likely to be observed with some radial velocity offset between the two components 
(expected to be a few tens km s$^{-1}$ for binaries with separations of a few AU). 
Additionally, repeat observations of close, bound pairs, such as those discovered by 
high-resolution imaging of nearby {\it K2} and {\it TESS} exoplanet host stars, will 
exhibit changes in the companion star RV over time. Stellar parameters can thus be 
compared at different epochs and so determined more reliably. Accordingly, it is 
important to carry out follow-up, high-resolution imaging observations to search 
for close companions and vet planet candidates. Furthermore, with 
sufficiently high spectral resolution, the spectral lines of a close, relatively bright 
companion star that are shifted by tens of km s$^{-1}$ could be detected, thus alerting 
to the presence of a spectroscopic binary.

\section{Conclusions}
\label{summ}

In this work we have quantified how the stellar parameters are affected when they 
are derived from blended spectra. We have explored the contribution of a companion 
star to a blended spectrum by varying the type of the star (i.e., various differences in 
$T_{\mathrm{eff}}$ between primary and secondary star) and its RV shift with respect to
the primary.
Typically, deviations in stellar parameters are up to 2-3 $\sigma$ from the values derived
from unblended spectra, with the effective temperatures and surface gravities least 
affected. Even stellar radii are not severely affected, given that their uncertainty (for 
unblended stars) is already of the order of 17\% when derived with {\tt SpecMatch-Emp}. 
The exceptions are relatively bright companion stars that are almost as bright as and 
also similar in $T_{\mathrm{eff}}$ to the primary star; these stars can cause an 
overestimation of the stellar radius by up to $\sim$ 60\%. We find that metallicities 
are very underestimated for all but the blends with the faintest companion stars; 
the metallicity is underestimated by an average of $\sim$ 4$\sigma$. 
These results are caused by the RV-shifted spectrum of the companion star 
added to the spectrum of the primary star, which alters some absorption lines that 
are used as indicators for $T_{\mathrm{eff}}$,  $\log (g)$, and [Fe/H]. In addition, 
the modified line profiles could affect the precision of RV measurements used for 
the determination of planet masses.
The $\chi^2$ of the fit is usually large when stellar parameters are significantly 
over- or underestimated, but there is not a linear correspondence between the 
$\chi^2$ value and the reliability of all parameters from a certain fit.

To account for the presence of a possible companion star, spectral fitting codes such
as {\tt SpecMatch-Emp} could be modified to include a second star in the fit. As was 
done in \citet{kolbl15}, after fitting the spectrum of the primary star, a search for a
second set of absorption lines could be performed on the residuals. However, this
method works only for secondary stars that are sufficiently bright (even though 
\citealt{kolbl15} claimed to be able to detect companions down to 1\% of the total flux) 
and with a sufficient RV shift ($\sim$ 10 km s$^{-1}$) with respect to the 
primary star. Since it relies on subtracting the spectrum of the primary star and finding
the secondary star in the residual spectrum, it is difficult to accurately determine the 
relative brightness of the secondary star and stellar parameters for secondaries
that are either similar in spectral type or brightness to the primary star or are 
very faint \citep{kolbl15}. Thus, ideally, spectral fitting codes could include an option
to fit two stars simultaneously, even though it might become prohibitively expensive
to carry out the computations (since the parameter space to explore is very large),
and it might not always be clear whether a single or a binary star fit yields
better results. More input from high-resolution images would be needed to detect 
or place robust  limits on possible companions that should be considered by the 
stellar fits.

To mitigate the effect of a contaminating star, obtaining a spectrum at different epochs
(and thus different RV shifts between the two stars if dealing with a bound system) 
would yield more realistic uncertainties for the stellar parameters of the primary star. 
However, this is only feasible for binaries with periods less than a few decades, 
and it would also require sufficient spectral resolution to separate the lines of the two 
stars (especially once the RV shifts reach just a few km s$^{-1}$).

The only way to identify close, bound companion stars is by obtaining follow-up observations: 
high-resolution imaging to detect nearby, faint companions, and high-resolution 
spectroscopy over a sufficient time baseline to detect any spectral lines belonging 
to a companion. Of course not all companions can be found, but particular attention
should be paid to bright companions, which affect stellar fits the most. 
Given that planet parameters sensitively depend on stellar parameters, we should 
aim at determining the most accurate stellar parameters possible, in particular if
we want to identify and characterize small, possibly habitable planets.

\acknowledgments
\begin{small}
We thank Johanna Teske and David Ciardi for constructive feedback which improved 
the paper. We also thank the referee whose comments made the discussion
presented in this paper clearer and more thorough.
Support for this work was provided by NASA through the NASA Exoplanet Exploration
Program Office. 
This research has made use of NASA's Astrophysics Data System Bibliographic Services.
It has also made use of the NASA Exoplanet Archive, which is operated by the California 
Institute of Technology, under contract with NASA under the Exoplanet Exploration 
Program.
\end{small}

\end{document}